\documentclass[10pt, twoside]{article}
\usepackage[utf8]{inputenc}
\usepackage[T1]{fontenc}
\usepackage[hypertexnames=false]{hyperref}
\usepackage{bm}

\usepackage[pdftex]{graphicx}
\usepackage{amsmath}
\usepackage{mathrsfs}
\usepackage{amsfonts}
\usepackage{amssymb}
\usepackage{mathtools}
\usepackage{array}
\usepackage{fancybox}
\usepackage{bm}
\usepackage{upgreek}
\usepackage{xcolor}
\usepackage{todonotes}
\usepackage{lineno}
\usepackage{comment}
\usepackage{makecell}
\usepackage{cancel}
\usepackage{empheq}
\usepackage{setspace} %doublespacing
\usepackage{nicefrac}

\usepackage{makecell}
\usepackage{multicol}

\usepackage[letterpaper,twoside,textwidth=18cm,textheight=25cm]{geometry}

\usepackage{biblatex}
\newcommand{\blat}[1]{\ensuremath{\mathbf{#1}}}
\newcommand{\bgre}[1]{\ensuremath{\bm{#1}}}

\newcommand{\boldb}{\blat{b}}

\newcommand{\bolde}{\blat{e}}

\newcommand{\boldf}{\blat{f}}

\newcommand{\boldL}{\blat{L}}
\newcommand{\boldm}{\blat{m}}
\newcommand{\boldM}{\blat{M}}
\newcommand{\boldn}{\blat{n}}

\newcommand{\boldr}{\blat{r}}

\newcommand{\boldt}{\blat{t}}
\newcommand{\boldu}{\blat{u}}

\newcommand{\x}{\blat{x}}
\newcommand{\X}{\blat{X}}
\newcommand{\y}{\blat{y}}

\newcommand{\boldz}{\blat{z}}

\newcommand{\boldtheta}{\bgre{\uptheta}}
\newcommand{\boldvarphi}{\bgre{\upvarphi}}
\newcommand{\boldomega}{\bgre{\upomega}}

\newcommand{\boldvarepsilon}{\bgre{\upvarepsilon}}
\newcommand{\boldchi}{\bgre{\upchi}}

\newcommand{\boldgamma}{\bgre{\upgamma}}
\newcommand{\boldkappa}{\bgre{\upkappa}}
\newcommand{\boldsigma}{\bgre{\upsigma}}

\newcommand{\boldmu}{\bgre{\upmu}}

\newcommand{\VRVE}{\ensuremath{V_0}}

\newcommand{\dd}[1]{\mathrm{\,d}\hspace{0.05em}#1}
\newcommand{\boldlevicivita}{\ensuremath{\bm{\mathcal{E}}}}
\newcommand{\levicivita}{\ensuremath{\mathcal{E}}}

\newcommand{\footremember}[2]{%
	\footnote{#2}
	\newcounter{#1}
	\setcounter{#1}{\value{footnote}}%
}

\begin{document}
%\doublespacing
\pagestyle{plain}
\pagenumbering{arabic}
%\linenumbers

\title{Do Discrete Fine-Scale Mechanical Models with Rotational Degrees of Freedom Homogenize Into a~Cosserat or a~Cauchy Continuum?}
\author{Jan Eliáš\footremember{Brno}{Brno University of Technology, Faculty of Civil Engineering, Brno, Czechia}\footremember{email}{Corresponding author: jan.elias@vut.cz} \and Gianluca Cusatis\footremember{Northwestern}{Northwestern University, Department of Civil and Environmental Engineering, Evanston, IL USA}}
\date{}

{\let\newpage\relax\maketitle}
\maketitle 

\section*{Abstract}
This article answers the question of whether homogenization of discrete fine-scale mechanical models, such as particle or lattice models, gives rise to an~equivalent continuum that is of Cauchy-type or Cosserat-type. The study employs the machinery of asymptotic expansion homogenization to analyze discrete mechanical models with rotational degrees of freedom commonly used to simulate the mechanical behavior of heterogeneous solids. The proposed derivation has general validity in both stationary (steady-state) and transient conditions  (assuming wavelength much larger that particle size)  and for arbitrary nonlinear, inelastic fine-scale constitutive equations. The results show that the unit cell problem is always stationary, and the only inertia term appears in the linear momentum balance equation at the coarse scale. Depending on the magnitude of the local bending stiffness, mathematical homogenization rigorously identifies two limiting conditions that correspond to the Cauchy continuum and the Cosserat continuum. An~heuristic combination of these two limiting conditions provides very accurate results also in the transition from one limiting case to the other. Finally, the study demonstrates that cases for which the Cosserat character of the homogenized response is significant are associated with non-physically high fine-scale bending stiffness and, as such, are of no interest in practice. 

\section*{Keywords}
homogenization, rotation, heterogeneity, discrete model, Cosserat continuum, Cauchy continuum, coarse scale, fine scale, length scale

%\tableofcontents

\section{Nomenclature}

\begin{multicols}{2}
\begin{description}
\itemsep -0.5em 
\item[$A$] area of contact
\item[$\boldb$] volume force, external load
\item[$d$] particle diameter
\item[$\bolde$] strain vector at interparticle contact
\item[$E_0$] elastic material constant
\item[$l$] length of contact $\equiv$ distance between particle centers
\item[$L_c$] structural size
\item[$\boldL$] first moment of inertia
\item[$\ell_c$] Cosserat length
\item[$\boldm$] couple traction vector at interparticle contact
\item[$\boldM$] second moment of inertia
\item[$\boldn_{\alpha}$] unit vector of local reference system
\item[$\boldr_{C\!I}$] vector pointing from particle $I$ to point $C$ 
\item[$\boldt$] traction vector at interparticle contact
\item[$\x_I$] coordinates of particle $I$
\item[$\x_{I\!J}$] branch vector from particle $I$ to $J$
\item[$\boldz$] volume couple, external load
\end{description}

\begin{description}
\itemsep -0.5em 
\item[$\alpha$] elastic material constant, local ref. system direction
\item[$\beta$] elastic material constant determining bending stiffness
\item[$\boldgamma$] strain tensor of Cosserat continuum
\item[$\varepsilon_V$] volumetric strain
\item[$\boldvarepsilon$] symmetric strain tensor of Cauchy continuum
\item[$\boldlevicivita$] Levi-Civita permutation tensor
\item[$\eta$] separation of scale
\item[$\boldtheta$] vector of total rotations 
\item[$\boldkappa$] curvature tensor
\item[$\boldmu$] couple stress tensor
\item[$\rho$] density
\item[$\boldsigma$] stress tensor
\item[$\boldvarphi$] independent part of rotation
\item[$\boldchi$] curvature vector at interparticle contact
\item[$\boldomega$] part of rotation dependent on translation
\end{description}
\end{multicols}

\section{Introduction}

Many engineering materials are heterogeneous on certain length scales depending upon the characteristics of their internal structures and the size of their major heterogeneity. Classical heterogeneous materials include geomaterials, concrete, wood, and composites, just to name a~few. A~more recent class of heterogeneous materials is that formed by metamaterials that are specifically engineered to exploit certain aspects of the internal structure heterogeneity to obtain unusual and unique engineering properties not found in traditional materials. The most effective and accurate approach to simulate the mechanical behavior of heterogenous materials is to simulate directly the complex system of interconnected major heterogeneities, e.g., stiff inclusions in a~more compliant matrix, via the interaction of simple components such as rigid particles connected with springs or truss and beams systems.  The main objective of this article presents general homogenization approach to discrete disordered systems with \emph{rotational degrees of freedom}, allowing arbitrary \emph{inelastic} material models and \emph{transient states} under assumption of wavelength much larger than size of material heterogeneities. 

Historically, discrete models have been formulated as either \emph{lattice} models or \emph{particle} models. The advantages of discrete models over traditional continuum-based formulations are summarized in the review paper~\parencite{BolEli-21}. Lattice models are systems of truss or beam elements. In mechanics, they were introduced by~\textcite{Hre-41} to solve elasticity problems. They were later adopted extensively to simulate fracture in quasi-brittle materials ~\parencite{SchMie92,JivEng-13,LukSav-16,ManMie08,ManMie11,MunRag-13}. In recent times, lattice models have been widely employed to simulate metamaterials ~\parencite{Zhu2022,Liu2015,Plesha2021,Kadic2016}.

Particle models are instead formulated via the interaction of rigid particles and with the definition of appropriate contact laws. The most popular particle model is arguably the so-called discrete element method (DEM) originally formulated for granular materials~\parencite{Cun71,CunStr78,CunStr79} and later adopted to simulate quasi-brittle materials like concrete \parencite{DonMag-99}. Particle models that have, contrary to DEM, a~fixed topology are also quite common. These models are often implemented via two-node links connecting adjacent particles, and, for this reason, they are sometimes characterized as lattice models or lattice-particle models. Among the most popular lattice-particle models, the pioneering paper of~\textcite{BolSai98}, based on the work of~\textcite{Kaw78}, proposed the Rigid-Body-Spring Network (RBSN) or Voronoi-Cell Lattice Model (VCLM) that has been used extensively~\parencite{AthWhe-18,AldHav-22,JosKun-23,YinJia-24,ShoNak-24,ParCho-24,EliSta12,BolXu-25}. 

Today, the most appreciated discrete model for quasi-brittle granular materials (e.g., concrete, rock) is the Lattice Discrete Particle Model (LDPM), which is directly connected to the heterogeneous features of the internal structure of the quasi-brittle material. Based on the early work on the Confinement-Shear Lattice (CSL) model~\parencite{CusCed07,CusBaz-03I,EliVor-15,EliVor20,CusBaz-03I}, LDPM made its first appearance in Refs.~\parencite{CusPel-11,CusMen-11} and has been used for a~number of different mechanical and multi-physical problems, including those dealing with complex loading scenarios~\parencite{PatTon-23,LyuPat-23,JiaBri-24}, failure of reinforced concrete~\parencite{GomBha-21,TroLal-25} and fiber reinforced concrete~\parencite{SchCus12,JinBur-16}, coupled multi-physics~\parencite{PatZha-23,YinTro-24,ZhuJia-25} and many other situations. 

In terms of kinematic and balance equations, there is no theoretical difference among all discrete models available in the literature. Usually, the main differences are in the constitutive formulation. One common feature of all these modes is that they feature rotational degrees of freedom that are independent of the displacement degrees of freedom. Hence, a~fundamental question is whether these models predict macroscopic behavior that is consistent with a~Cauchy continuum (as assumed in Ref.~\parencite{SanMun25}) or a~Cosserat continuum. This study explores this question by adopting the concepts of asymptotic homogenization theory.

The theory of asymptotic homogenization allows one to explore the macroscopic properties of heterogeneous materials. Developed in 1980s~\parencite{BenLio-78,San80}, asymptotic homogenization allows separating material response at different length scales by looking at the limit where scale separation becomes infinitely large. It develops kinematic, constitutive, and balance equations at various length scales and provide the link among them. Homogenization techniques are often used only to study the macroscopic constitutive behavior, but the upscaling of kinematic and balance equations is equally important and enlightening. Analytical solutions exist only for very few cases of the homogenization problem; in most situations, one must rely on approximate numerical solutions. Therefore, the computational homogenization or FE$^2$ approach has become the standard homogenization tool~\parencite{SmiBre-98,AlaGan-21,LebPon-21}. %Nowadays, the homogenized systems are built with fine scale models replaced by machine learning techniques~\parencite{CorKal-25,TanGhn-25,FlaSte-25}.

It is well-known that heterogeneous systems exhibit wave dispersion and frequency band gaps. To find an equivalent homogeneous continuum with similar dispersion properties, dynamic homogenization and various continuation methods have been developed~\parencite{TorFan-25,GaoHe-25,TorBel-23,PerShm18} resulting in enriched higher order continua~\parencite{BacBad-25,TorFan-24,CusZho14}. Even though rotational degrees of freedom are often considered, these methods are usually applied to periodic discrete systems with elastic material models. In contrast to this work, they are focused on wavelengths comparable to the size of the heterogeneities. The dispersion also occurs on the structural scale due to structural geometry or a combination of materials with different mechanical properties~\parencite{MuhLim-19}. However, such a phenomenon has no relation to internal material heterogeneity and is therefore not considered relevant for the homogenization scheme considered here.

Inspired by the work of \textcite{FisChe-07} dealing with homogenization of atomistic systems, \textcite{RezCus16} derived the asymptotic homogenization of the mechanical behavior of discrete model with rotational degrees of freedom. The same homogenization methodology was later applied in Refs.~\parencite{RezAln-19,RezZho-17}. They concluded that the corresponding coarse scale model is a~Cosserat continuum, but they also verified the Cosserat effects to be negligible for LDPM. 
From the viewpoint of the asymptotic homogenization theory, the derivation in \textcite{RezCus16} suffered from allowing the Representative Volume Element (RVE) to rotate, which is not consistent with the periodicity assumption adopted to separate the problem at multiple length scales.   

\textcite{EliCus22} homogenized the coupled problem of mass transport and mechanics in a~discrete setting. In this work, the RVE rotation was restricted, but a~higher order angular momentum balance equation was used at the coarse scale. Consequently, the coarse-scale Cosserat continuum was obtained again, and once again the Cosserat effects were verified to be negligible for LDPM. It is worth noting that the transport portion of the derivation, published separately in Ref.~\parencite{EliYin-22} as well as the developed coupling terms previously published are not affected by the present derivation.

The present study corrects previous inconsistencies and significantly extends the published conclusions. It is inspired by the %~excellent
paper of~\textcite{ForPra-01}, which presented the asymptotic homogenization of the heterogeneous Cosserat continuum.  The dissertation of Francis Pradel~\parencite{PradelThesis} extends this work also to discrete systems. Unfortunately, it is published in French only and had not been known to the authors at the time they worked on the present article. In steady-state linear elasticity, the present article and Pradel's thesis delivers identical results. Nevertheless, there are several differences: (i) the \emph{inertia terms} are included, and consequently, the homogenization incorporates transient terms; (ii) the homogenization is derived for arbitrary fine-scale \emph{nonlinear inelastic} constitutive models; (iii) the rotations are considered to be the sum of an~\emph{independent} component, that is asymptotically expanded, and \emph{dependent} component, which is given by the curl of the displacement field. The last difference directly introduces the symmetric and antisymmetric displacement gradients, which were missing in the original paper~\parencite{ForPra-01} and dissertation~\parencite{PradelThesis}. The present work also delivers extensive numerical verification of the derived equations in steady-state, transient, and also inelastic regimes.

\section{Fundamentals of discrete mechanical models}

The \emph{primary} variables in the mechanical discrete models~\parencite{BolEli-21} are kinematic degrees of freedom: a~displacement vector, $\boldu$, and a~rotation vector, $\boldtheta$. Both $\boldu$ and $\boldtheta$ have three components in a~three-dimensional space.

\begin{figure}[!tb]
\centering
\includegraphics[width=12cm]{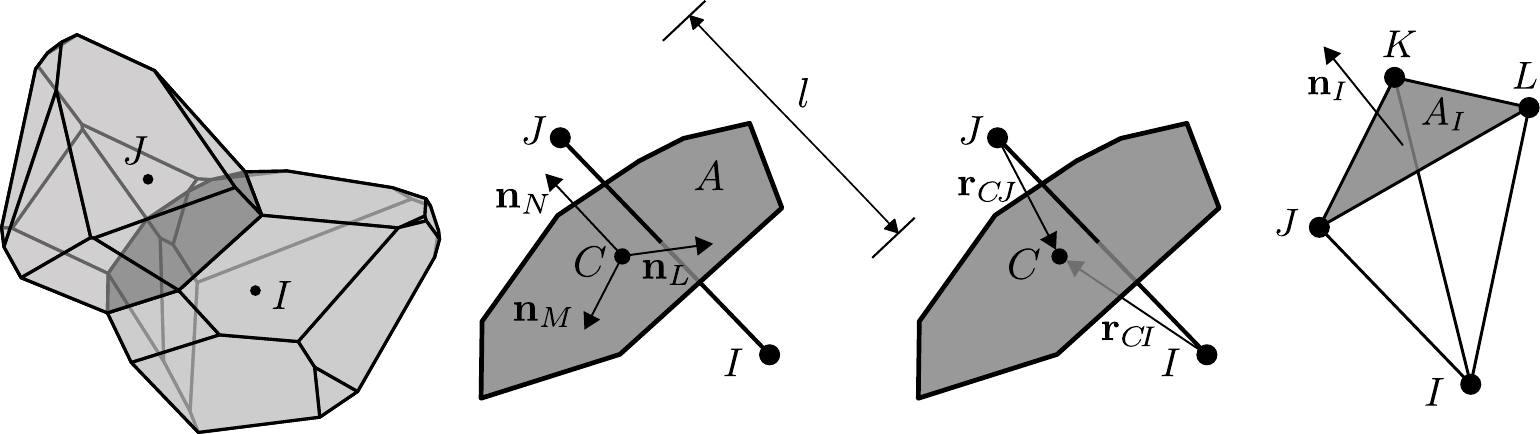}
\caption{Two rigid bodies $I$ and $J$ in contact; local reference system and vectors $\boldr_{C\!I}$ and $\boldr_{C\!J}$; tetrahedron for evaluation of volumetric strain.} \label{fig:facet}
\end{figure}

The \emph{intermediate} variables consist of the strain vector, $\bolde$, and the curvature vector, $\boldchi$. Both are vectors with three components in three dimensions. They are defined by the following \emph{kinematic} equations~\parencite{BolEli-21} valid under the assumption of small rotations and small displacement at any point $C$ at coordinates $\x_C$ at the contact facet (Fig. \ref{fig:facet}) between nodes $I$ and $J$
\begin{subequations} \label{eq:disstraincurv}
\begin{align}
 e_{\alpha} &= \frac{1}{l}\left[\boldu_{J} - \boldu_{I} + \boldlevicivita : \left(\boldtheta_{J}\otimes \boldr_{C\!J} - \boldtheta_{I}\otimes \boldr_{C\!I} \right) \right]\cdot\boldn_{\alpha} \label{eq:disstrain}\\
\chi_{\alpha} &= \frac{1}{l}\left[\boldtheta_J- \boldtheta_I \right]\cdot\boldn_{\alpha} \label{eq:discurv}
\end{align}
\end{subequations}
where $\boldr_{C\!I}$ (or $\boldr_{C\!J}$) is a~vector connecting particle governing node $I$ (or $J$, respectively) with the point of interest $C$ on the contact face, $\boldlevicivita$ is the Levi-Civita permutation tensor and $\boldn_{\alpha}$ are local normal and two tangential directions ($\alpha\in\left\{N,\,M,\,L\right\}$), see Fig.~\ref{fig:facet}. The length of the connecting strut is $l=||\x_{I\! J}||$ and the contact normal direction is $\boldn_{N}=\x_{I\! J}/l$ where  $\x_{I\! J} = \x_J-\x_I$. 

The normal direction of the contact $\boldn_N$ can in general be different from the true facet normal $\boldn_F$. This is, for example, the case of LDPM or other models that are based on a~different type of tessellation other than Voronoi or power tessellations, see, e.g., Ref.~\parencite{Eli20}. To account for this misalignment, the contact area $A$ is projected orthogonal to the connecting direction: $A^{\star}=A \boldn_N\cdot\boldn_F$. Also, in different models, the shape of the area $A$ can be different, e.g. a~triangle for LDPM and a~polygon for RBSM (VCLM). This difference does not affect in any way the derivation presented in this paper. 

In addition to the fundamental kinematic equations~\eqref{eq:disstraincurv}, for certain constitutive equations that capture confinement, one also needs to calculate the volumetric strain, $\varepsilon_V$. This can be calculated as one third of the volume change associated with a~generic facet. In LDPM, for example, since the nodal connectivity is based on a~Delaunay tetrahedralization, the volumetric strain can be calculated with reference to each LDPM tetrahedron. In this case, under the assumption of small displacements, the volumetric strain is linearly dependent on $\boldu$, and reads~\cite{EliCus22}
\begin{align}
\varepsilon_V &= -\frac{1}{9V_t} \sum\limits_{I\in t} A_I \boldu_I \cdot \boldn_I \label{eq:volumetric_strain}
\end{align}
where $I$ runs over the four vertices of the tetrahedron $t$ with volume $V_t$; $\boldu_I$ is the displacement vector of node $I$, and $A_I$ and $\boldn_I$ are the area and outward normal of the triangular face of the tetrahedron opposite to node $I$. For models other than LDPM, there might be several tetrahedrons associated with one contact facet, in that case the volumetric strain is simply the average of the volumetric strains from all the attached tetrahedrons.

The second set of governing equations is the \emph{constitutive} equations, which relate strain and curvature to the \emph{flux} variables, specifically the traction vector, $\boldt$, and the couple traction vector, $\boldm$. This study assumes that these constitutive equations are, apart from thermodynamic admissibility, arbitrary and can be written in the following general form
\begin{align}
\boldt &= \boldf_s\left(\bolde,\,\varepsilon_V\right) & \boldm &= \boldf_m\left(\boldchi\right)  \label{eq:distconstitutive_eq}
\end{align}
The adopted constitutive equations assume that the couple traction is dependent solely on curvature and the traction is dependent on both the strain vector and volumetric strain. The effect of volumetric strain is included to account for the effect of confinement~\parencite{CusPel-11,CusMen-11}. One can easily imagine other variables entering these functions such as temperature, pressure, and various other physical quantities appearing in coupled multi-physical simulations as well as history dependent variables. 

The final set of governing equations enforces the \emph{balance} of linear and angular momentum to each rigid particle $I$
\begin{subequations} \label{eq:LDPMbalance}
\begin{align}
\rho V_I \ddot{\boldu}_I + \rho \boldL_I \cdot \ddot{\boldtheta}_I - V_I \boldb &= \sum_{e\in I} A^{\star}_e t_{\alpha e}\boldn_{\alpha e} 
\\
\rho \boldM_I \cdot \ddot{\boldtheta}_I + \rho \boldL_I^T \cdot \ddot{\boldu}_I - V_I \left(\boldz + \boldlevicivita:\boldr_{0I}\otimes\boldb\right) &= \sum_{e\in I} A^{\star}_e \left[t_{\alpha e}\boldlevicivita:(\boldr_{C\!I}\otimes \boldn_{\alpha e}) + m_{\alpha e}\boldn_{\alpha e}\right]
\end{align}
\end{subequations}
where $e$ runs over all contacts of the particle $I$ of volume $V_I$ and density $\rho$. The index $e$ referring to all contacts of the particle $I$ is added to individual contact variables when it is meaningful. The summation over local directions $\alpha$ is not explicitly written by the summation symbol; instead, the Einstein summation rule is adopted. The summations on the right-hand sides integrate the tractions into forces and the couple tractions and the moment of tractions into couples by multiplying them with the contact area; $\boldL_I$ and $\boldM_I$ are the first and second moments of inertia tensors of the particle $I$ defined as
\begin{align} \label{eq:inertia}
\boldM_I & = \int_{V_I}  (\boldr_I \cdot \boldr_I\, \bm{1}  - \boldr_I \otimes \boldr_I) \dd{V_I} &
\boldL_I = -\boldL_I^T &= V_I \boldlevicivita\cdot \boldr_{0I}
\end{align}
where $\boldr_I$ is a~vector pointing from the particle governing node $\x_I$ to any point within the particle and $\boldr_{0I}$ is a~vector pointing from $\x_I$ to the particle centroid; $\bm{1}$ is the second-order identity tensor. For the special case $\boldr_{0}=\bm{0}$ where the governing node of the particle coincides with its centroid, $\boldL=\bm{0}$. This situation occurs in certain particular cases, e.g., if one uses a~centroidal Voronoi tessellation to define the particle geometry, but not in general. Finally, $\boldb$ represents an~external volume force, and $\boldz$ represents an~external volume couple, both of which are vectors with three components. The expression $\boldlevicivita:\boldr_{0I}\otimes\boldb$ accounts for the couple due to the volume force load acting at eccentricity $\boldr_0$, that is, the difference between particle centroid and particle governing node.

\section{Separation of scales}
Even if the material is not periodic, for homogenization purposes, this study assumes that there exists a~periodic fine-scale structure known as the Representative Volume Element (RVE) with characteristic size $d_c$, which is the first length scale of the problem. It is worth noting that here the RVE is considered ``representative'' only with reference to the geometrical description of the internal structure and not necessarily to its mechanical behavior. The material heterogeneity is assumed to repeat periodically with respect to the RVE size. The second fundamental length scale is the structural length scale $L_c$, which can be characterized by the maximum distance between two points in the coarse-scale domain. The coarse scale lives in a~global reference system $\X$. Its dimensionless version, $\tilde \X$, is obtained by normalizing the global coordinates by $L_c$. Similarly, dimensionless local coordinates in the RVE local reference systems $\y$ can be defined at each point of the coarse-scale domain by means of the fine-scale size $d_c$. One has 
\begin{align}
\X &= L_c\tilde \X & \y &= d_c \tilde \y 
\end{align}
This implies that the scales of $\X$ and $\y$ are related via a~positive separation of scale constant $\eta=d_c/L_c$. 

The asymptotic expansion homogenization studies a~limiting situation when $\eta$ approaches zero, $\eta\rightarrow 0$. Each dimensionless quantity $\tilde \bullet=\tilde \bullet(\tilde\X,\tilde\y)$ appears now as a~function of both the global and local normalized reference systems $\tilde \X$ and $\tilde \y$ and can be approximated by infinite series with terms at different scales~\parencite{BenLio-78,San80}. Starting with the displacement, one can write 
\begin{align} 
\tilde\boldu(\tilde\X,\tilde\y) &= \tilde\boldu^{(0)}(\tilde\X,\tilde\y) +  \eta\tilde\boldu^{(1)}(\tilde\X,\tilde\y) + \eta^2\tilde\boldu^{(2)}(\tilde\X,\tilde\y) + \dots  \label{eq:expansion_u}
\end{align}
All components of each expansion must be periodic in $\tilde\y$ with period 1 and must have, except the first, zero mean in the $y$ domain. The scale on which the displacements reside is assumed to be equal to the structural scale: $u_c\sim L_c \sim \mathcal{O}(\eta^0)$. 

The rotations $\boldtheta$ are not expanded directly as is done by~\textcite{ForPra-01} for the homogenization of heterogeneous Cosserat continua and in Refs.~\parencite{PradelThesis,RezCus16,EliCus22} homogenizing discrete models. Instead, as first proposed in \textcite{RezCus16} for the homogenization of discrete models, the rotation must be divided into two additive components, one that \emph{depends} on displacements and another one that is \emph{independent}
\begin{align} 
\tilde\boldtheta = \tilde\boldomega + \tilde\boldvarphi \label{eq:rot_decomp}
\end{align}
The dependent part, $\boldomega$, is dictated by the displacement field. In the continuum case, it would be half the curl of the displacement field: $\boldomega=\nicefrac{1}{2}\,\nabla\times\boldu = \nicefrac{1}{2}\,\boldlevicivita:\nabla\otimes\boldu$. It is straightforward to show that the contraction of the Levi-Civita permutation tensor with the continuous dependent rotation becomes the antisymmetric part of the displacement gradient
\begin{align}
\boldlevicivita\cdot\boldomega =  \frac{1}{2}\boldlevicivita\cdot\boldlevicivita:\nabla\otimes\boldu = \frac{1}{2}\left(\nabla\otimes\boldu - \boldu\otimes\nabla\right) = \nabla\stackrel{\mathrm{a}}\otimes\boldu \label{eq:antisymmetric}
\end{align}
The spatial differentiation operator, $\nabla$, can be expressed according to the chain rule as $\nabla\rightarrow \nabla_{X} + \nabla_{y} = L_c^{-1}\nabla_{\tilde X} + d_c^{-1}\nabla_{\tilde y}$. The dependent rotation then reads
\begin{align}
\boldomega &= \frac{1}{2} \boldlevicivita:\left( \nabla_{X} \otimes \boldu + \nabla_{y} \otimes \boldu\right) &\Leftrightarrow&&   \omega_c\tilde\boldomega =  \frac{1}{2}\boldlevicivita:\left(\eta\nabla_{\tilde X}\otimes\tilde\boldu + \nabla_{\tilde y}\otimes\tilde\boldu \right)\frac{u_c}{d_c} \label{eq:deprot}
\end{align}
where on the right-hand side the equation is conveniently expressed with separated dimensionless and dimensional terms showing the scale of the dependent rotation to be $\omega_c\sim u_c/d_c \sim L_c/d_c \sim \mathcal{O}(\eta^{-1})$.

In this study, the asymptotic expansion is applied only to the \emph{independent} part of the rotation, $\tilde\boldvarphi$,  since the dependent one is already expanded through displacements. Of course, it is always possible to combine the expanded independent rotation, $\tilde \boldvarphi$, with the dependent rotation from the expanded displacement, $\tilde \boldu$, to obtain the expansion of the full rotation, $\tilde\boldtheta$.
\begin{align} 
\tilde\boldvarphi(\tilde\X,\tilde\y) &= \tilde\boldvarphi^{(-1)}(\tilde\X,\tilde\y) +\eta\tilde\boldvarphi^{(0)}(\tilde\X,\tilde\y) + \eta^2\tilde\boldvarphi^{(-1)}(\tilde\X,\tilde\y) + \dots \label{eq:expansion_phi}
\end{align}
The scale of the independent rotation is assumed to be identical to the scale of dependent rotations. Therefore, $\varphi_c\sim\omega_c\sim\mathcal{O}(\eta^{-1})$ and the dimensionless expansion begins with the term $\tilde\boldvarphi^{(-1)}$ to emphasize that rotations are one order below displacements. This notation is not standard in classical homogenization texts, but it clearly elucidates the resulting structure of the separated equations.  The upper index in the first expanded term will therefore always indicate the scale of the original variable hereinafter.

From the point of view of the spatial coordinates at the coarse scale $\tilde\X$, the nodes $I$ and $J$ are close to each other. According to \textcite{FisChe-07}, the coarse-scale gradient $\nabla_{\tilde X}$ at coordinates $(\tilde \X_I,\,\tilde \y_J)$ can be used to approximate the mechanical field variables at coordinates $(\tilde \X_J,\,\tilde \y_J)$. 
\begin{subequations} \label{eq:Taylorseries}
\begin{align}
\tilde\boldu(\tilde \X_J,\tilde \y_J) &= \tilde\boldu(\tilde\X_I,\tilde\y_J) + \eta\frac{\partial \tilde\boldu(\tilde\X_I,\tilde\y_J)}{\partial \tilde X_j} \tilde x^{I\! J}_j +  \eta^2\frac{1}{2}\frac{\partial^2 \tilde\boldu(\tilde\X_I,\tilde\y_J)}{\partial \tilde X_j \tilde X_k} \tilde x^{I\! J}_j \tilde x^{I\! J}_k + \dots \label{eq:Taylorseries_u}\\
\tilde\boldvarphi(\tilde \X_J,\tilde \y_J) &= \tilde\boldvarphi(\tilde\X_I,\tilde\y_J) + \eta\frac{\partial \tilde\boldvarphi(\tilde\X_I,\tilde\y_J)}{\partial \tilde X_j} \tilde x^{I\! J}_j +  \eta^2\frac{1}{2}\frac{\partial^2 \tilde\boldvarphi(\tilde\X_I,\tilde\y_J)}{\partial \tilde X_j \tilde X_k} \tilde x^{I\! J}_j \tilde x^{I\! J}_k + \dots
\end{align} 
\end{subequations}
It is critical here to apply the Taylor expansion only to the independent part of the rotation. The dependent part is already expanded with displacements in Eq.~\eqref{eq:Taylorseries_u}. These equations allow us to treat the discrete system at the coarse scale as continuous. Since the right-hand side always uses the same $\tilde\X$ coordinate, the notation of the primary variables will be simplified hereinafter to $\bullet^{(\psi I)} = \bullet^{(\psi)}(\X,\y_I)$ for the indicial notation and $\bullet^{(\psi)}_I$ for the tensorial notation. For example, the $i$th component of displacement $u^{(0)}_i(\X,\y_I)$ will be written as $u^{(0I)}_i$ or $(\boldu^{(0)}_I)_i$.

One can finally also write an~expansion of the dependent part of the rotation. Equations~\eqref{eq:deprot} and \eqref{eq:expansion_u} yield
\begin{align} 
\tilde\boldomega{(\tilde\X,\tilde\y_I)} &= \underbrace{\frac{1}{2}\boldlevicivita:\nabla_{\tilde y}\otimes\tilde \boldu^{(0)}_I}_{\boldomega^{(-1I)}}  + \eta\underbrace{\frac{1}{2}\boldlevicivita:\left(\nabla_{\tilde X}\otimes\tilde \boldu^{(0)}_I + \nabla_{\tilde y}\otimes\tilde \boldu^{(1)}_I \right)}_{\boldomega^{(0I)}} + \eta^2\underbrace{\frac{1}{2}\boldlevicivita:\left(\nabla_{\tilde X}\otimes\tilde \boldu^{(1)}_I + \nabla_{\tilde y}\otimes\tilde \boldu^{(2)}_I \right)}_{\boldomega^{(1I)}}  + \dots   \label{eq:expansion_omega}
\end{align}
To conclude the asymptotic expansion of rotations, the normalized total rotations $\tilde\boldtheta(\tilde\X,\tilde\y_I) = \tilde \boldtheta^{(-1)}_I + \eta \tilde \boldtheta^{(-1)}_I + \dots$ according to Eq.~\eqref{eq:rot_decomp} are reported.
\begin{subequations}
\label{eq:theta_expansion}
\begin{align}
\tilde\boldtheta^{(-1)}_I &=\tilde\boldvarphi^{(-1)}_I+\tilde\boldomega^{(-1)}_I = \tilde\boldvarphi^{(-1)}_I + \frac{1}{2}\boldlevicivita:\nabla_{\tilde y}\otimes\tilde\boldu^{(0)}_I \label{eq:theta-1}
\\
\tilde\boldtheta^{(0)}_I &= \tilde\boldvarphi^{(0)}_I + \tilde\boldomega^{(0)}_I = 
\tilde\boldvarphi^{(0)}_I + \frac{1}{2}\boldlevicivita:\left( \nabla_{\tilde y} \otimes \tilde\boldu^{(1)}_I  +  \nabla_{\tilde X} \otimes \tilde\boldu^{(0)}_I\right) \label{eq:theta0}
\\
\tilde\boldtheta^{(1)}_I &= \tilde\boldvarphi^{(1)}_I + \tilde\boldomega^{(1)}_I =
\tilde\boldvarphi^{(1)}_I + \frac{1}{2}\boldlevicivita:\left( \nabla_{\tilde y} \otimes \tilde\boldu^{(2)}_I  +  \nabla_{\tilde X} \otimes \tilde\boldu^{(1)}_I\right) \label{eq:theta1}
\end{align}
\end{subequations}

Combining the expansions of the primary variables~\eqref{eq:expansion_u} and \eqref{eq:expansion_phi} with Taylor expansion \eqref{eq:Taylorseries}, and inserting them together with Eqs.~\eqref{eq:expansion_omega} and \eqref{eq:antisymmetric} into the kinematic equation~\eqref{eq:disstraincurv} yields an~asymptotic expansion of strain and curvature. The dimensionless strains and curvatures of contact $e$ between rigid bodies $I$ and $J$ read
\begin{align}
\tilde\bolde &= \tilde\bolde^{(-1)} + \eta\tilde\bolde^{(0)} +  \eta^2\tilde\bolde^{(1)} + \dots &
\tilde\boldchi &= \tilde\boldchi^{(-2)} + \eta\tilde\boldchi^{(-1)} + \eta^2\tilde\boldchi^{(0)} + \dots
\label{eq:strain_expansion}
\end{align}
where the individual components are listed in Appendix~\ref{sec:strain_expansion}, $\tilde\bolde=\bolde/e_c$, and $\tilde\boldchi$=$\boldchi/\chi_c$. Based on Eq.~\eqref{eq:disstraincurv}, one can deduce that the strains must be of order $e_c \sim \mathcal{O}(\eta^{-1})$, and the curvatures must be of order $\chi_c \sim \mathcal{O}(\eta^{-2})$, which is reflected in the first terms appearing in Eq.~\eqref{eq:strain_expansion}.
Note that in this step, the antisymmetric part of the displacement gradient comes into play due to the splitting of the rotation into the \emph{dependent} and \emph{independent} part.

Similarly, the asymptotic expansion of the volumetric strain reads
\begin{align}
\tilde\varepsilon_V &= \tilde\varepsilon_V^{(-1)} + \eta\tilde\varepsilon_V^{(0)} +  \eta^2\tilde\varepsilon_V^{(1)} + \dots \label{eq:volumetric_strain_expansion}
\end{align}
where $\tilde\varepsilon_V = \varepsilon_V/e_c$, and the individual terms are again listed in Appendix~\ref{sec:strain_expansion}. The normalization constant for volumetric strain is obviously $e_c$, the same as the one used to normalize the strain vector.

The constitutive model maps strains and curvatures to tractions, $\boldt$, and couple tractions, $\boldm$. The scale of tractions, $t_c$ should be proportional to the scale of strains, $e_c$, multiplied by the scale of the elastic modulus, $E_c$ (assumed $E_c\sim\mathcal{O}(\eta^0)$). The normalization constant for couple traction is, according to Refs.~\parencite{ForPra-01,PradelThesis}, proportional to $E_c$ and the squared Cosserat characteristic length scale $\ell_c^2$ with $\ell_c$ considered proportional to $\mathcal{O}(\eta^0)$. Consequently, $t_c\sim\mathcal{O}(\eta^{-1})$, $m_c\sim\mathcal{O}(\eta^{-2})$, and dimensionless expansions read
\begin{align} 
\tilde\boldt &= \tilde\boldt^{(-1)} +  \eta\tilde\boldt^{(0)} + \dots  &
\tilde\boldm &= \tilde\boldm^{(-2)} +  \eta\tilde\boldm^{(-1)} + \dots 
\end{align}
In order to distinguish two limiting cases, the dimensionless asymptotic expansion of the Cosserat length is introduced
\begin{align} 
\tilde\ell(\tilde\X,\tilde\y) &= \tilde\ell^{(0)}(\tilde\X,\tilde\y) +  \eta\tilde\ell^{(1)}(\tilde\X,\tilde\y) + \eta^2\tilde\ell^{(2)}(\tilde\X,\tilde\y) + \dots  \label{eq:expansion_ell}
\end{align}
In the first limiting case LC1 (corresponding to hypothesis HS2 in Refs.~\parencite{ForPra-01,PradelThesis}), the first term in the Cosserat length expansion is non-zero. Therefore, the total Cosserat length is proportional to the structural size $L_c$ and the first terms of constitutive equations read
\begin{align} 
\tilde\boldt^{(-1)} &= \tilde\boldf_s \left(\tilde\bolde^{(-1)},\,\tilde\varepsilon_V^{(-1)}\right)  & \tilde\boldm^{(-2)} &=  \tilde\boldf_m \left(\tilde\boldchi^{(-2)}\right) \label{eq:const0_LC1}
\end{align}
$\tilde\boldf_s$ and $\tilde\boldf_m$ are dimensionless forms of the constitutive model~\eqref{eq:distconstitutive_eq}.

In the second limiting case LC2 (corresponding to hypothesis HS1 in Refs.~\parencite{ForPra-01,PradelThesis}), the first term of Cosserat length expansion is assumed to vanish, $\tilde\ell^{(0)}=0$, that is, the Cosserat length is proportional to the fine-scale size $d_c$. The \emph{effective} squared Cosserat length scale then becomes $\mathcal{O}(\eta^2)$ and the first two terms in the expansion of couple tractions vanish: $\tilde\boldm^{(-2)}=\tilde\boldm^{(-1)}=\bm{0}$. 
\begin{align} 
\tilde\boldt^{(-1)} &= \tilde\boldf_s \left(\tilde\bolde^{(-1)},\,\tilde\varepsilon_V^{(-1)}\right)  & \tilde\boldm^{(0)} &=  \tilde\boldf_m \left(\tilde\boldchi^{(-2)}\right) \label{eq:const0_LC2}
\end{align}
If either $\tilde\bolde^{(-1)}$ (and $\tilde\varepsilon_V^{(-1)}$) or $\tilde\boldchi^{(-2)}$ are zero (as will be found later), the constitutive equations must be evaluated with reference to the first non-zero value of strains and curvatures. Higher terms, if needed, shall be determined by Taylor expansion.

Similarly to the primary variables, the notation of the intermediate variables (strain and curvature) and flux variables (traction and couple traction) can be simplified as follows. For the contact $e$ between the nodes $I$ and $J$, the $\alpha$ component of the intermediate or flux variable $\bullet$ will be written as $(\bullet^{(\psi)}_e)_{\alpha} = \bullet^{(\psi e)}_{\alpha}$. For example, the normal component of the traction $\boldt^{(-1)}$ of contact $e$ is $(\boldt^{(-1)}_e)_N$  or $t^{(-1e)}_N$.

Finally, one needs to develop a~dimensionless version of the balance equations. Rewriting Eq.~\eqref{eq:LDPMbalance} gives
\begin{subequations}
\label{eq:balanceX}
\begin{align}
\frac{\rho_c d_c^3 \tilde\rho}{\tau_c^2}\left(u_c
\tilde V_I
\frac{\partial^2 \tilde \boldu_I}{\partial \tilde \tau^2} + \theta_c d_c \tilde \boldL_I \cdot \frac{\partial^2 \tilde \boldtheta_I}{\partial \tilde \tau^2}\right)  - d_c^3 b_c \tilde V_I \tilde \boldb &= d_c^2 t_c \sum_{e\in I} \tilde A^{\star}_e \tilde t_{\alpha e}\boldn_{\alpha e} 
\\
\frac{\rho_c d_c^4 \tilde\rho}{\tau_c^2}\left(\theta_c d_c\tilde\boldM_I \cdot \frac{\partial^2 \tilde \boldtheta_I}{\partial \tilde \tau^2} + u_c\tilde\boldL_I^T \cdot \frac{\partial^2 \tilde \boldu_I}{\partial \tilde \tau^2}\right)  - d_c^3 \tilde V_I \left(z_c  \tilde\boldz  + d_c  b_c \boldlevicivita:\tilde\boldr_{0I}\otimes \tilde\boldb\right) &= d_c^2 \sum_{e\in I} \tilde A_e^{\star}\left[d_c t_c \tilde t_{\alpha e}\boldlevicivita:(\tilde \boldr_{C\!I}\otimes \boldn_{\alpha e}) + m_c \tilde m_{\alpha e}\boldn_{\alpha e}\right]
\end{align}
\end{subequations}
where all the variables are transformed into their dimensionless versions and $\rho_c$ is the normalization constant for the density $\rho$. Note that the second time derivatives, $\ddot\bullet$, are now also performed with respect to dimensionless time $\tilde\tau = \tau/\tau_c$, and therefore the double dots are no longer used. Volume and area are scaled as $V=d_c^3\tilde V$ and $\tilde A^{\star}=d_c^2\tilde A^{\star}$ since their length scale is dictated by the size of the particles, which is on the same length scale as the fine-scale size $d_c$. Based on Eq.~\eqref{eq:inertia} the transformations of the inertia tensors of the particles read $\boldM=d_c^5 \tilde \boldM$ and $\boldL=d_c^4 \tilde \boldL$, respectively. In addition, the vector $\boldr = d_c \tilde \boldr$ also scales with the length scale of the particles.

All previously developed scaling constants are now substituted into the balance equations~\eqref{eq:balanceX}. In order to obtain the same units as typically used in continuum mechanics (that is N/m$^3$ for the linear and Nm/m$^3$ for the angular momentum balance), both equations are divided by the volume of the particles $V_I=d_c^3 \tilde V_I$.
\begin{subequations} 
\begin{align}
\frac{\rho_c u_c \tilde\rho}{\tau_c^2 \tilde V_I}\left(\tilde V_I \frac{\partial^2 \tilde \boldu_I}{\partial \tilde \tau^2} + \tilde \boldL_I \cdot \frac{\partial^2 \tilde \boldtheta_I}{\partial \tilde \tau^2}\right) - b_c \tilde \boldb &= \frac{E_c u_c}{d_c^2 \tilde V_I} \sum_{e\in I} \tilde A^{\star}_e\tilde t_{\alpha e} \boldn_{\alpha e} 
\\
\frac{\rho_c u_c d_c \tilde \rho}{\tau_c^2 \tilde V_I}\left(\tilde\boldM_I \cdot \frac{\partial^2 \tilde \boldtheta_I}{\partial \tilde \tau^2} + \tilde\boldL_I^T \cdot \frac{\partial^2 \tilde \boldu_I}{\partial \tilde \tau^2}\right) - z_c  \tilde\boldz  - b_c d_c \boldlevicivita:\tilde\boldr_{0I}\otimes \tilde\boldb &= \frac{E_c u_c}{d_c \tilde V_I} \sum_{e\in I} \tilde A^{\star}_e\left[\tilde t_{\alpha e} \boldlevicivita:(\tilde \boldr_{C\!I}\otimes \boldn_{\alpha e}) + \frac{\ell_c^2}{d^2_c} \tilde m_{\alpha e}\boldn_{\alpha e}\right]
\end{align}
\label{eq:normalized-balance-eqs}
\end{subequations}

Next, it is reasonable to assume that the body forces and the body couples are both of order zero, $b_c \sim z_c \sim \mathcal{O}(\eta^{0})$. Similarly to \textcite{chen2006generalized}, this study also assumes that the mass density is of order zero: $\rho_c \sim \mathcal{O}(\eta^{0})$. It is worth noting that these assumptions implies that elastic wave speed $\sqrt{E_c/\rho_c}$ also resides at the coarse scale, and consequently, the dynamic wavelength of the response is of the order of zero if only one time scale is considered: $\tau_c\sqrt{E_c/\rho_c} \sim \mathcal{O}(\eta^{0})$. This assumption will result in transferring all the translational time-dependent effects to the coarse scale. It is valid only for dynamic responses characterized by small frequency content, which is usually sufficient for simulating civil engineering structures loaded by earthquake or live loads. The dispersion of waves exhibited by heterogeneous systems for large frequencies will be lost by this choice. Other options capable to represent behavior of the short wavelengths are also possible; for example, see homogenization approaches presented in Refs.~\parencite{AurBou-09,chen2006generalized} or continuation techniques~\parencite{TorBel-23} specifically designed for wavelength comparable to particle sizes.

Based on previous discussion, one has $\rho_c u_c / \tau^2_c \sim \mathcal{O}(\eta^{0})$, $\rho_c u_c d_c/ \tau^2_c \sim \mathcal{O}(\eta^{1})$, $E_c u_c / d_c^2 \sim \mathcal{O}(\eta^{-2})$, and $E_c u_c / d_c \sim \mathcal{O}(\eta^{-1})$. The last constant appearing in the balance equations is $\ell_c^2/d_c^2$. Based on the previous consideration, this constant is proportional to $\eta^{-2}$. However, in the limiting case LC2 (Cosserat length proportional to the fine-scale size), the first two couple tractions are zero. 

In conclusion, the balance equations in their asymptotic dimensionless form read
\begin{subequations}\label{eq:balance_dimensionless}
\begin{align}
\tilde \rho\frac{\partial^2 \left(\tilde \boldu^{(0)}_I + \eta\tilde \boldu^{(1)}_I + \dots\right)}{\partial \tilde \tau^2} + \frac{\tilde \rho \tilde \boldL_I}{\tilde V_I} \cdot \frac{\partial^2 \left(\tilde \boldtheta^{(-1)}_I + \eta \boldtheta^{(0)}_I + \dots\right)}{\partial \tilde \tau^2} - \tilde \boldb = \frac{1}{\eta^2\tilde V_I}\sum_{e\in I} \tilde A_I^{\star} \left(\tilde t^{(-1e)}_{\alpha}+\eta\tilde t^{(0e)}_{\alpha}+\dots\right)\boldn_{\alpha e} 
\\
\eta\frac{\tilde \rho\tilde\boldM_I}{\tilde V_I} \cdot \frac{\partial^2 \left(\tilde \boldtheta^{(-1)}_I+\tilde \boldtheta^{(0)}_I+\dots\right)}{\partial \tilde \tau^2} +\eta\frac{\tilde \rho \tilde\boldL_I^T}{\tilde V_I} \cdot \frac{\partial^2 \left(\tilde \boldu^{(0)}_I+\tilde \boldu^{(1)}_I+\dots\right)}{\partial \tilde \tau^2} -  \tilde\boldz - \eta\boldlevicivita:\tilde\boldr_{0I}\otimes\tilde\boldb= \nonumber \\
\dfrac{1}{\eta \tilde V_I}\displaystyle\sum_{e\in I} \tilde A_e^{\star}\left[\left(\tilde t^{(-1e)}_{\alpha} + \eta\tilde t^{(0e)}_{\alpha} + \dots\right)\boldlevicivita:(\tilde \boldr_{C\!I}\otimes \boldn_{\alpha e}) + \eta^{-2} \left(\tilde m^{(-2e)}_{\alpha}+\eta\tilde m^{(-1e)}_{\alpha}+\dots\right)\boldn_{\alpha e}\right]
\end{align}
\end{subequations}
These equations separate all terms into appropriate scales depending on the power of $\eta$. Scaled back from the dimensionless versions and multiplied by volume $V_I$, the linear momentum equations at the lowest scales read
\begin{subequations} \label{eq:balance_LC1_linear}
\begin{align}
&\eta^{-2}:&\bm{0} &= \sum_{e\in I} A_e^{\star} t^{(-1e)}_{\alpha}\boldn_{\alpha e}   \label{eq:LC1-t-2}
\\
&\eta^{-1}:&\bm{0} &= \sum_{e\in I} A_e^{\star} t^{(0e)}_{\alpha}\boldn_{\alpha e}   \label{eq:LC1-t-1}
\\
&\eta^{0}:& \rho V_I \ddot{\boldu}^{(0)}_I + \rho\boldL_I \cdot \ddot{\boldtheta}^{(-1)}_I - V_I \boldb &= \sum_{e\in I} A_e^{\star} t^{(1e)}_{\alpha}\boldn_{\alpha e}  \label{eq:LC1-t0}
\end{align}
\end{subequations}
The relevant angular momentum balances read
\begin{subequations} \label{eq:balance_LC1_angular}
\begin{align}
&\eta^{-3}:& \bm{0} &= \sum_{e\in I} A_e^{\star} m^{(-2e)}_{\alpha}\boldn_{\alpha e} \label{eq:LC1-m-3}
\\
&\eta^{-2}:& \bm{0} &= \sum_{e\in I} A_e^{\star}  m^{(-1e)}_{\alpha}\boldn_{\alpha e} \label{eq:LC1-m-2}
\\
&\eta^{-1}:& \bm{0} &=  \sum_{e\in I} A_e^{\star}\left[t^{(-1e)}_{\alpha}\boldlevicivita:(\boldr_{C\!I}\otimes \boldn_{\alpha e}) + m^{(0e)}_{\alpha}\boldn_{\alpha e}\right] \label{eq:LC1-m-1}
\\
&\eta^{0}:& -V_I\boldz &=  \sum_{e\in I} A_e^{\star}\left[ t^{(0e)}_{\alpha}\boldlevicivita:(\boldr_{C\!I}\otimes \boldn_{\alpha e}) + m^{(1e)}_{\alpha}\boldn_{\alpha e}\right] \label{eq:LC1-m0}
\end{align}
\end{subequations}
The next two sections present the first-order solutions to the developed multiscale problems.
%and are solved hierarchically scale by scale. (i) The lowest scales should provide constantness of the coarse-scale variables across the fine-scale, (ii)  the next scale defines the fine-scale problem. (iii) Finally, at the next scale, one looks at the balance of the entire fine-scale model to obtain the coarse-scale problem definition. The overall fine-scale model balance in the case of the linear momentum means sums all the equations for individual particles. However, for the angular momentum, one needs to rewrite all the balances with respect to some reference point. The easiest option is to use the (arbitrary) origin of the reference system $y$; therefore, all the $\boldr$ vectors that point to the particle governing node $\y_I$ are rewritten to $\y_I + \boldr$.

%Traditionally in the asymptotic analysis of a~continuum, the first problem appears at order $\eta^{-2}$ since the lowest flow variables (stress in mechanics) reside at order $\eta^{-1}$ and the balance equation features its $y$-derivative. Consequently, problems (ii) and (iii) appear on the scale $\eta^{-1}$ and $\eta^{0}$, respectively. This is also valid here for the limiting case LC1 and linear momentum balances in the limiting case LC2. 

%However, the LC2 angular momentum balances reside at different scales.

%%%%%%%%%%%%%%%%%%%%%%%%%%%%%%%%%%%%%%%%%%%%%%%%%%%%%%%%%%%%%%%%%%%%%%%%
\section{Limiting case  LC1: Cosserat length proportional to structural size}
%%%%%%%%%%%%%%%%%%%%%%%%%%%%%%%%%%%%%%%%%%%%%%%%%%%%%%%%%%%%%%%%%%%%%%%%
The first limiting case assumes that the first term of the Cosserat length expansion~\eqref{eq:expansion_ell} is non-zero, $\ell^{(0)}\neq 0$.
 
%%%%%%%%%%%%%%%%%%%%%%%%%%%%%%%%%%%%%%%%%%%%%%%%%%%%%%%%%%%%%%%%%%%%%%%%
\subsection{LC1: The \texorpdfstring{$\mathcal{O}(\eta^{-3})$}{O(-3)} problem}
The lowest scale contains the balance of the angular momentum \eqref{eq:LC1-m-3} as a~function of the $\mathcal{O}(\eta^{-2})$ couple. The moments of traction do not appear in the equation because asymptotically they are negligible compared to the couples.
The constitutive equation is taken from Eq.~\eqref{eq:const0_LC1}.
\begin{subequations}
\begin{align}
\bm{0} &= \sum_{e\in I} A_e^{\star} m^{(-2e)}_{\alpha}\boldn_{\alpha e} 
&
\boldm^{(-2)} &=  \boldf_m \left(\boldchi^{(-2)}\right)
\end{align}
\end{subequations}
The corresponding kinematic equation is Eq.~\eqref{eq:curv-2}. The primary kinematic variables are $\boldu^{(0)}$ and $\boldvarphi^{(-1)}$, but the kinematic equation can be simplified using the total rotation $\boldtheta^{(-1)}$ from Eq.~\eqref{eq:theta-1}. Equation~\eqref{eq:curv-2} then reads
\begin{align}
\chi^{(-2)}_{\alpha} &= \frac{\boldn_{\alpha}}{l}\cdot\left(\boldtheta^{(-1)}_J - \boldtheta^{(-1)}_I\right) &
\end{align}
The only solution in the linear regime that satisfies the $y$-periodicity is the $y$-constant total rotation $\boldtheta^{(-1)}$, zero curvature $\boldchi^{(-2)}$ and zero couple traction $\boldm^{(-2)}$. 

In the inelastic regime, there might be other solutions.  However, the elastic solution presented here is still valid and a~reasonable choice also for the nonlinear case. The $y$-constantness of the coarse-scale primary variables represents a~solution with the lowest potential energy; therefore, it should be maintained throughout the loading history. 

%%%%%%%%%%%%%%%%%%%%%%%%%%%%%%%%%%%%%%%%%%%%%%%%%%%%%%%%%%%%%%%%%%%%%%%%
\subsection{LC1: The \texorpdfstring{$\mathcal{O}(\eta^{-2})$}{O(-2)} problem}
The next scale consists of the balance of linear and angular momentum \eqref{eq:LC1-t-2} and \eqref{eq:LC1-m-2}
\begin{align}
\bm{0} &= \sum_{e\in I} A_e^{\star} t^{(-1e)}_{\alpha}\boldn_{\alpha e} 
&
\bm{0} &= \sum_{e\in I} A_e^{\star} m^{(-1e)}_{\alpha}\boldn_{\alpha e} 
\end{align}
Both constitutive relations \eqref{eq:const0_LC1} are used. However, notice that since the curvature $\boldchi^{-2}$ is zero, the constitutive equation for couple traction is now expressed with reference to the curvatures $\mathcal{O}(\eta^{-1})$. 

\begin{align}
\boldt^{(-1)} &= \boldf_s \left(\bolde^{(-1)},\,\varepsilon_V^{(-1)}\right) 
&
\boldm^{(-1)} &=  \boldf_m \left(\boldchi^{(-1)}\right)
\end{align}
Finally, the kinematic equations \eqref{eq:str-1} and \eqref{eq:curv-1} together with the definition of volumetric strain $\varepsilon_V^{(-1)}$ from Eq.~\eqref{eq:ev-1} appear on this scale. They can be rewritten using the total rotation $\boldtheta^{(0)}$ and $\boldtheta^{(-1)}$ from Eqs.~\eqref{eq:theta_expansion} as 
\begin{subequations}
\begin{align}
\bolde^{(-1)}_{\alpha} &=
\frac{\boldn_{\alpha}}{l}\cdot\left[\boldu^{(0)}_J - \boldu^{(0)}_I +\boldlevicivita:\left(\boldtheta^{(-1)}_J\otimes\boldr_{C\!J} - \boldtheta^{(-1)}_I\otimes\boldr_{C\!I} \right)\right] 
\\
\boldchi^{(-1)}_{\alpha} &= \frac{\boldn_{\alpha}}{l}\cdot\left( \boldtheta^{(0)}_J - \boldtheta^{(0)}_I +   \x_{I\!J} \cdot \nabla_{X}\otimes\boldtheta^{(-1)}_J\right) 
\\
\varepsilon^{(-1)}_V &= -\frac{1}{9 V_t} \sum\limits_{I\in t} A_I \boldu^{(0)}_I \cdot \boldn_I
\end{align}
\end{subequations}

The only solution that satisfies the $y$-periodicity is the $y$-constant displacement $\boldu^{(0)}$ and rotation $\boldtheta^{(0)}$, and zero rotation $\boldtheta^{(-1)}=\bm{0}$. The strain $\bolde^{(-1)}$, as well as the curvature $\boldchi^{(-1)}$, the traction $\boldt^{(-1)}$, and the couple traction $\boldm^{(-1)}$ are also zero. Moreover, the volumetric strain $\varepsilon_V^{(-1)}$ from Eq.~\eqref{eq:ev-1} also becomes zero for $y$-constant $\boldu^{(0)}$.
\begin{align}
\varepsilon^{(-1)}_V &= -\frac{1}{9 V_t} \boldu^{(0)}_I \cdot \sum\limits_{I\in t} A_I  \boldn_I = -\frac{1}{9 V_t} \boldu^{(0)}_I \cdot \int\limits_{\Gamma_t} \boldn \dd{\Gamma_t}  = -\frac{1}{9 V_t} \boldu^{(0)}_I \cdot \int\limits_{V_t} \nabla_{x}\cdot\bm{1} \dd{ V_t} = \bm{0} \label{eq:zero_ev0}
\end{align}
The above equation exploits the divergence theorem to transfer integration from the tetrahedral surface to its volume.

In case of inelastic material behavior, other solutions might exist. Using the same arguments used for the scale $\mathcal{O}(\eta^{-3})$, it is possible and reasonable to consider the linear solution to be also valid for inelastic constitutive models.  

%%%%%%%%%%%%%%%%%%%%%%%%%%%%%%%%%%%%%%%%%%%%%%%%%%%%%%%%%%%%%%%%%%%%%%%%
\subsection{LC1: The \texorpdfstring{$\mathcal{O}(\eta^{-1})$}{O(-1)} problem}
The balance equations on the next higher scale, \eqref{eq:LC1-t-1} and \eqref{eq:LC1-m-1}, must be solved numerically. Because traction $\boldt^{(-1)}$ is zero, these equations read
\begin{align}
\bm{0} &= \sum_{e\in I} A_e^{\star} t^{(0e)}_{\alpha}\boldn_{\alpha e}
&
\bm{0} &=  \sum_{e\in I} A_e^{\star} m^{(0e)}_{\alpha}\boldn_{\alpha e}
\label{eq:LC1-1_balances}
\end{align}
The associated constitutive equation must be evaluated with reference to the $\mathcal{O}(\eta^{0})$ strains and curvatures since the higher-order strains and curvatures are all equal to zero. 

\begin{align}
\boldt^{(0)} &= \boldf_s \left(\bolde^{(0)},\,\varepsilon_V^{(0)}\right) 
&
\boldm^{(0)} &=  \boldf_m \left(\boldchi^{(0)}\right) \label{eq:LC1-1_constitutive}
\end{align}
The kinematic equations are Eqs.~\eqref{eq:str0}, \eqref{eq:curv0}, and \eqref{eq:ev0X}. Using again the total rotations from Eq.~\eqref{eq:theta_expansion} along with derived $y$-costantness of $\boldu^{(0)}$ and $\boldtheta^{(0)}$ and zero $\boldtheta^{(-1)}$, the set of kinematic equations become
\begin{subequations}
\begin{align}
 \bolde^{(0)}_{\alpha} &= 
\frac{\boldn_{\alpha}}{l}\cdot\left[ \boldu^{(1)}_J -  \boldu^{(1)}_I +  \x_{I\!J}\cdot \nabla_{X}\otimes  \boldu^{(0)}  + \boldlevicivita: \boldtheta^{(0)}\otimes\left(\boldr_{C\!J} - \boldr_{C\!I} \right)\right] \label{eq:str0_LC1}
\\
 \boldchi^{(0)}_{\alpha} &= \frac{\boldn_{\alpha}}{l}\cdot\left( \boldtheta^{(1)}_J -  \boldtheta^{(1)}_I +
 \x_{I\!J}\cdot\nabla_{X}\otimes \boldtheta^{(0)} \right) 
\\
\varepsilon^{(0)}_V &= -\frac{1}{9  V_t} \left(\sum\limits_{I\in t}  A_I  \boldu^{(1)}_I \cdot \boldn_I\right) + \frac{1}{3} \mathrm{tr} \left( \nabla_{ X}  \cdot \boldu^{(0)} \right) 
\end{align}
\end{subequations}
With identity $\boldr_{C\!I}-\boldr_{C\!J}=\x_{I\!J}$ substituted into Eq.~\eqref{eq:str0_LC1}, they can be rewritten as functions of primary variables on this scale and a~loading component in the form of eigen-strains and eigen-curvature projected from strains and curvatures from the lower scale.
\begin{subequations}
\label{eq:LC1_eigen}
\begin{align}
 e^{(0)}_{\alpha} &= \frac{1}{l}\left(\boldu^{(1)}_J - \boldu^{(1))}_I \right) \cdot \boldn_{\alpha} - \hat e_{\alpha} &\text{where}\quad \hat e_{\alpha} &= -\boldn_N \cdot \left(\nabla_X \otimes \boldu^{(0)} - \boldlevicivita\cdot\boldtheta^{(0)}\right)\cdot\boldn_{\alpha} = -\boldn_N \cdot \boldgamma\cdot\boldn_{\alpha}
\\
\chi^{(0)}_{\alpha} &= \frac{1}{l}\left(\boldtheta^{(0)}_J - \boldtheta^{(0)}_I\right) \cdot \boldn_{\alpha} - \hat\chi_{\alpha} &\text{where}\quad \hat\chi_{\alpha} &=  -\boldn_N \cdot \left(\nabla_X \otimes \boldtheta^{(0)} \right)\cdot\boldn_{\alpha} = -\boldn_N \cdot \boldkappa\cdot\boldn_{\alpha}
\\
\varepsilon^{(0)}_V &= -\frac{1}{9  V_t} \left(\sum\limits_{I\in t} A_I  \boldu^{(1)}_I \cdot \boldn_I\right) - \hat\varepsilon_V &\text{where}\quad \hat\varepsilon_V &= -\frac{1}{3}\mathrm{tr}\left(\boldgamma\right)
\end{align}
\end{subequations}

The derived steady-state fine-scale discrete problem is solved numerically. The degrees of freedom are the fine-scale displacement $\boldu^{(1)}$ and the total fine-scale rotation $\boldtheta^{(1)}$. The limiting balances of linear and angular momentum are completely decoupled, the displacement problem has no interaction with the rotation problem, and vice versa. This is because asymptotically (i) moments due to tractions $\boldt^{(0)}$ are on a~higher scale compared to couple tractions $\boldm^{(0)}$ and (ii) translations of contact points due to rigid-body rotations $\boldtheta^{(1)}$ are on a~higher scale compared to rigid-body translations $\boldu^{(1)}$. The decoupled fine-scale problem can be solved by implementing a~special discrete element with altered kinematics and statics given by Eqs.~\eqref{eq:LC1_eigen} and \eqref{eq:LC1-1_balances}, that is the rotation removed from the kinematic equation and moment of traction is removed from the balance equation.

The loading is provided by the coarse-scale kinematic variables introduced as eigen-strain and eigen-curvature. These eigen-strains and  eigen-curvatures are calculated as projections of the coarse-scale Cosserat strain $\boldgamma=\nabla_X \otimes \boldu^{(0)} - \boldlevicivita\cdot\boldtheta^{(0)}$ and curvature $\boldkappa = \nabla_X \otimes \boldtheta^{(0)}$, the volumetric eigen-strain is the negative coarse-scale volumetric strain $\mathrm{tr}(\boldgamma)/3$. 

The boundary conditions for both problems are given by the assumed periodicity of the displacement and rotation fields. Yet another boundary condition is needed since only differences of the primary unknowns are involved and therefore there exist an~infinite number of solutions differing by a~constant. The additional boundary conditions state that, on average over fine-scale model of volume $\VRVE$, both displacement $\boldu^{(1)}$ and rotation $\boldtheta^{(1)}$ should be zero, because they represent fluctuations added to the main trends $\boldu^{(0)}$ and $\boldtheta^{(0)}$. 
\begin{align} \label{eq:LC1_bc}
\left\langle\boldu^{(1)}\right\rangle&= \frac{1}{\VRVE} \sum_{I\in\VRVE} V_I \boldu^{(1)}_I = \bm{0} & \left\langle\boldtheta^{(1)}\right\rangle & = \frac{1}{\VRVE} \sum_{I\in\VRVE} V_I \boldtheta^{(1)}_I = \bm{0}  
\end{align}   
These conditions can be enforced as a~linear constraint or via the Lagrange multiplier. Advantageously, one can also easily impose them during post-processing by shifting the whole solution. The first approach unfortunately leads to a~great increase of computational cost (as discussed in Ref.~\parencite{EliYin-22}).

%%%%%%%%%%%%%%%%%%%%%%%%%%%%%%%%%%%%%%%%%%%%%%%%%%%%%%%%%%%%%%%%%%%%%%%%
\subsection{LC1: The \texorpdfstring{$\mathcal{O}(\eta^{0})$}{O(0)}problem}

Once the fluctuation fields on the $\mathcal{O}(\eta^{-1})$ are resolved, the focus shifts to the overall balance of the fine-scale problem. The equations for linear momentum balances~\eqref{eq:balance_LC1_linear} and angular momentum balances \eqref{eq:balance_LC1_angular} are now considered summed over the whole fine-scale model. All the previously considered equations at lower scales are automatically satisfied in their summed form, since they were already resolved locally. The balance equations on this scale are~\eqref{eq:LC1-t0} and \eqref{eq:LC1-m0}, in the summed version they read
\begin{subequations}
\label{eq:LC1mac_summed}
\begin{align}
\left\langle\rho\right\rangle \ddot{\boldu}^{(0)}  - \left\langle\boldb\right\rangle &= \frac{1}{\VRVE}\sum_{I\in\VRVE} \sum_{e\in I} A_e^{\star} t^{(1e)}_{\alpha} \boldn_{\alpha e} \label{eq:LC1mac_linsummed}
\\
- \left\langle\boldz\right\rangle &= \frac{1}{\VRVE}\sum_{I\in\VRVE} \sum_{e\in I} A_e^{\star}\left[t^{(0e)}_{\alpha}\boldlevicivita:\left(\y_I+\boldr_{C\!I}\right)\otimes \boldn_{\alpha e} + m^{(1e)}_{\alpha}\boldn_{\alpha e}\right] \label{eq:LC1mac_angsummed}
\end{align}
\end{subequations}
The balance of angular momentum is summed with respect to some arbitrary reference point chosen here as the origin of the reference system $y$, therefore, the vectors $\boldr_{C\!I}$ have been changed to $\y_I+\boldr_{C\!I}$. Nevertheless, the symbols $\y_I$ can be immediately removed from the equation, since $\boldlevicivita:\y_I\otimes\left(\sum_{e\in I} A_e^{\star} t^{(0e)}_{\alpha}\boldn_{\alpha e}\right)$ must be zero for every rigid body according to the fine-scale balance~\eqref{eq:LC1-1_balances}.  

Both balance equations feature summation over all particles $I$ within the fine-scale model of volume $\VRVE$ and then summation over all contacts $e$ attached to the particle $I$ oriented from $I$ towards its neighboring particle $J$. Therefore, each contact is visited twice, once oriented as $I\!J$ and once as $J\!I$. These contacts have the opposite reference system, $^{I\!J}\!\boldn_{\alpha}=-^{J\!I}\!\boldn_{\alpha}$ but identical effective areas $A^{\star}$, lengths $l$, tractions $\boldt^{(0)}$, and couple tractions $\boldm^{(0)}$.

The balance equations feature traction $\boldt^{(1)}$ and couple traction $\boldm^{(1)}$, which are estimated by Taylor expansion from Eq.~\eqref{eq:LC1-1_constitutive} (originally Eq.~\ref{eq:const0_LC1})
\begin{subequations}
\label{eq:const_taylor_expansion}
\begin{align} 
t^{(1)}_{\alpha} & \approx e^{(1)}_{\beta} \frac{\partial \left(\boldf_s(\bolde, \varepsilon_V)\right)_{\alpha}}{\partial  e_{\beta}}\bigg|_{\substack{\bolde=\bolde^{(0)}\\\varepsilon_V=\varepsilon_V^{(0)}}}
= \frac{\partial t^{(0)}_{\alpha}} {\partial  e^{(0)}_{\beta}} e^{(1)}_{\beta} + \frac{\partial t^{(0)}_{\alpha}} {\partial  \varepsilon^{(0)}_V} \varepsilon^{(1)}_V 
&
m^{(1)}_{\alpha} & \approx \chi^{(1)}_{\beta} \frac{\partial \left(\boldf_m(\boldchi)\right)_{\alpha}}{\partial \chi_{\beta}}\bigg|_{\boldchi=\boldchi^{(0)}}
= \frac{\partial m^{(0)}_{\alpha}} {\partial \chi^{(0)}_{\beta}}\chi^{(1)}_{\beta}
\end{align}
\end{subequations}
These constitutive equations are substituted into the balances~\eqref{eq:LC1mac_summed}. Since the differentiations are identical and the normal direction is opposite, the double summation visiting each element twice with opposite orientation leads to a~difference in the strain $\bolde^{(1)}$, the curvature $\boldchi^{(1)}$, and the volumetric strain $\varepsilon_V^{(1)}$. The kinematic equations on this scale are Eqs.~\eqref{eq:str1}, \eqref{eq:curv1}, and \eqref{eq:ev1}.
Rather than rewriting them here, it is useful to directly derive the mentioned differences. Clearly, the volumetric strain in these opposite elements $\varepsilon_V^{(1)}$ is identical, as the volumetric strains in the associated tetrahedrons given by Eq.~\eqref{eq:ev1} are identical. The strains and curvatures differences, after careful rearrangement of the terms from Eqs.~\eqref{eq:str1}, \eqref{eq:curv1}, provide
\begin{align}
^{I\!J}\! e^{(1)}_{\alpha} - ^{J\!I}\!\! e^{(1)}_{\alpha} &= \x_{I\!J} \cdot \left(\nabla_{ X}  e^{(0)}_{\alpha} \right)
&
^{I\!J}\!\chi^{(1)}_{\alpha} - ^{J\!I}\!\!\chi^{(1)}_{\alpha} &= \x_{I\!J} \cdot \left(\nabla_{ X} \chi^{(0)}_{\alpha}\right)
&
^{I\!J}\!\varepsilon_V^{(1)} - ^{J\!I}\!\!\varepsilon_V^{(1)} &= 0 \label{eq:strainsum}
\end{align}
The summed balance equations~\eqref{eq:LC1mac_summed} then yield
\begin{subequations}
\begin{align}
\left\langle\rho\right\rangle \ddot{u}^{(0)}_i  - \left\langle b_i\right\rangle &= \frac{1}{\VRVE}\sum_{e\in\VRVE} A_e^{\star} l_e n^{N\!e}_j \frac{\partial t^{(0e)}_{\alpha}} {\partial  e^{(0e)}_{\beta}}  \frac{\partial e^{(0e)}_{\beta}} {\partial  X_j}  n^{\alpha e}_i
\\
- \left\langle z_i \right\rangle &= \frac{1}{\VRVE}\sum_{e\in\VRVE} A_e^{\star} \left[t^{(0e)}_{\alpha}\levicivita_{ijk}\left(r^{C\!I}_j - r^{C\!J}_j \right) n^{\alpha e}_k + l_e n^{N\!e}_j\frac{\partial m^{(0e)}_{\alpha}} {\partial \chi^{(0e)}_{\beta}} \frac{\partial \chi^{(0e)}_{\beta}} {\partial  X_j}  n^{\alpha e}_i\right] 
\end{align}
\end{subequations}

The chain rule of differentiation replaces the derivative with respect to $\bolde^{(0)}$ and $\boldchi^{(0)}$ by the derivative with respect to $\X$. In addition, the difference of the rigid arms is replaced $\boldr_{C\!I}-\boldr_{C\!J}=\x_{I\!J}$. Finally, the differentiation with respect to $\X$ can be extended over all expressions involving area, length, and local reference directions since these variables are constant with respect to $\X$.

The momentum balances can be written by defining the coarse-scale stress tensor, $\boldsigma$, and the couple stress tensor, $\boldmu$, as
\begin{subequations} \label{eq:macrobalance_LC1}
\begin{align}
\left\langle\rho\right\rangle \ddot{\boldu}^{(0)}  - \left\langle\boldb\right\rangle &= \nabla_{X}\cdot \boldsigma \label{eq:LC1_Cosserat_linear}
\\
- \left\langle\boldz\right\rangle &= \nabla_{X}\cdot \boldmu + \boldlevicivita:\boldsigma
\label{eq:LC1_Cosserat_angular}
\end{align}
\end{subequations}
where
\begin{subequations} \label{eq:macroflux_LC1}
\begin{align}
\boldsigma &= \frac{1}{\VRVE}\sum_{e\in\VRVE} l_e A_e^{\star} t^{(0e)}_{\alpha} \boldn_{N\!e} \otimes \boldn_{\alpha e} \label{eq:LC1_macrostress}
\\
\boldmu &= \frac{1}{\VRVE}\sum_{e\in\VRVE} l_e A_e^{\star} m^{(0e)}_{\alpha} \boldn_{N\!e} \otimes \boldn_{\alpha e} \label{eq:LC1_macrocouplestress}
\end{align}
\end{subequations}
These equations reveal the coarse-scale representation as a~continuous homogeneous Cosserat medium with degrees of freedom being the coarse-scale displacement, $\boldu^{(0)}$, and the total coarse-scale rotation, $\boldtheta^{(0)}$. The stress and couple stress variables are collected projections of traction and couple traction computed at the fine-scale level. There are two independent, decoupled and steady-state fine-scale models loaded by projection of the Cosserat strain $\boldgamma=\nabla_X \otimes \boldu^{(0)} - \boldlevicivita\cdot\boldtheta^{(0)}$, curvature $\boldkappa=\nabla_X \otimes \boldtheta^{(0)}$, and volumetric strain $\varepsilon_V=\mathrm{tr}(\boldgamma)/3$. The result obtained corresponds to the findings presented by~\textcite{PradelThesis} for steady-state linear elasticity under hypothesis HS2. 

Notice that the inertia terms related to rotational degrees of freedom completely vanished, the only transient term remaining is the one with the second time derivative of coarse-scale displacement. Unlike displacements in which the entire fine scale moves due to $\boldu^{(0)}$ as a~rigid body, the coarse-scale rotation $\boldtheta^{(0)}$ expresses constant rotations of the individual particles, which give asymptotically irrelevant inertia components. 

The famous expression for coarse-scale stress~\eqref{eq:LC1_macrostress} is usually attributed to~\textcite{Love1927} and~\textcite{Web66} and is called the Love-Weber formula. It has been derived in many publications~\parencite{RotSel81,ChrMeh-81,BalMar07,NicHad-13,LinWu16} and it also exactly corresponds to the formula derived previously in Ref.~\parencite[Eq.~22]{RezCus16} and Ref.~\parencite[Eq.~31b]{EliCus22} by asymptotic expansion homogenization. As argued in Ref.~\parencite{EliCus25} or \parencite{YanReg19}, the expression is not correct in a~general case where the load acts outside the particle governing nodes $\x_I$; the missing term is called \emph{boundary-radius gap} in the literature. The periodic model emerging from the homogenization does not have any external load, and the expression is therefore exact. 

The coarse-scale couple stress formula~\eqref{eq:LC1_macrostress} differs from expressions used in literature~\parencite{ChaLia-90,BarVar01,ChaKuh05,Vard18_book} or derived previously by the authors~\parencite{RezCus16,EliCus22,EliCus25}, it misses the contributions of couples due to tractions on eccentricity. This is because the derivation here is done under the strict assumption of asymptotically small $\eta$, which separates the moments due to traction to the higher scale compared to couple tractions.

%%%%%%%%%%%%%%%%%%%%%%%%%%%%%%%%%%%%%%%%%%%%%%%%%%%%%%%%%%%%%%%%%%%%%%%%
\section{Limiting case LC2: Cosserat length proportional to the fine-scale size}
%%%%%%%%%%%%%%%%%%%%%%%%%%%%%%%%%%%%%%%%%%%%%%%%%%%%%%%%%%%%%%%%%%%%%%%%
%%%%%%%%%%%%%%%%%%%%%%%%%%%%%%%%%%%%%%%%%%%%%%%%%%%%%%%%%%%%%%%%%%%%%%%%
The second limiting case considers that the first term in the expansion of the characteristic Cosserat length is zero: $\ell^{(0)}=0$. This implies that the first two couple tractions automatically vanish, $\boldm^{(-2)}=\boldm^{(-1)}=\bm{0}$, and the lowest curvature, $\boldchi^{(-2)}$, is not zero in general. It becomes the leading term for the calculation of $\boldm^{(0)}$ via the constitutive function~\eqref{eq:const0_LC2}.

The separated balance
equations~\eqref{eq:balance_LC1_linear} and \eqref{eq:balance_LC1_angular} from limiting case LC1 remain valid also for LC2. However, they cannot be solved sequentially because there is a~coupling of the problems on the various scales through the primary kinematic variables.

%%%%%%%%%%%%%%%%%%%%%%%%%%%%%%%%%%%%%%%%%%%%%%%%%%%%%%%%%%%%%%%%%%%%%%%%
\subsection{LC2: The \texorpdfstring{$\mathcal{O}(\eta^{-3})$}{O(-3)} problem}
The lowest scale is given by \eqref{eq:LC1-m-3}. However, since we now consider that the first nonzero term in the expansion of the Cosserat length~\eqref{eq:expansion_ell} is $\ell^{(1)}$, the first nonzero term in the expansion of couple tractions becomes $\boldm^{(0)}$. The balance equation~\eqref{eq:LC1-m-3} is, therefore, satisfied automatically.

%%%%%%%%%%%%%%%%%%%%%%%%%%%%%%%%%%%%%%%%%%%%%%%%%%%%%%%%%%%%%%%%%%%%%%%%
\subsection{LC2: The \texorpdfstring{$\mathcal{O}(\eta^{-2})$}{O(-2)} problem}
The balance equation~\eqref{eq:LC1-m-2} vanishes as the couple traction $\boldm^{(-1)}=0$. The remaining equations at this scale to solve are the linear momentum balance~\eqref{eq:LC1-t-2}, constitutive equation for tractions~\eqref{eq:const0_LC2}, and the kinematic equations~\eqref{eq:str-1} and \eqref{eq:ev-1}. Using the definition of total rotation from Eq.~\eqref{eq:theta-1}, these equations read
\begin{subequations} \label{eq:LC2_0-2}
\begin{align}
\bm{0} &= \sum_{e\in I} A_e^{\star} t^{(-1e)}_{\alpha}\boldn_{\alpha e} & \boldt^{(-1)} &= \boldf_s \left(\bolde^{(-1)},\,\varepsilon_V^{(-1)}\right)
\\ 
\bolde^{(-1)}_{\alpha} &=
\frac{\boldn_{\alpha}}{l}\cdot\left[\boldu^{(0)}_J - \boldu^{(0)}_I +\boldlevicivita:\left(\boldtheta^{(-1)}_J\otimes\boldr_{C\!J} - \boldtheta^{(-1)}_I\otimes\boldr_{C\!I} \right)\right]  & \varepsilon^{(-1)}_V &= -\frac{1}{9 V_t} \sum\limits_{I\in t}  A_I  \boldu^{(0)}_I \cdot \boldn_I  \label{eq:LC2_0-2_kinematics}
\end{align}
\end{subequations}

The degrees of freedom are translations $\boldu^{(0)}$ and rotations $\boldtheta^{(-1)}$. The solution should satisfy the $y$-periodicity. However, there are too many primary variables compared to the available number of equations. The needed additional equations appear on the next scale.  

%%%%%%%%%%%%%%%%%%%%%%%%%%%%%%%%%%%%%%%%%%%%%%%%%%%%%%%%%%%%%%%%%%%%%%%%
\subsection{LC2: The \texorpdfstring{$\mathcal{O}(\eta^{-1})$}{O(-1)} problem}
Let us start with the angular momentum balance equation~\eqref{eq:LC1-m-1}, associated constitutive model~\eqref{eq:const0_LC2}, and the kinematic equation~\eqref{eq:curv-2} rewritten with the help of definition~\eqref{eq:theta0}. 
\begin{align}
\bm{0} &=  \sum_{e\in I} A_e^{\star}\left[t^{(-1e)}_{\alpha}\boldlevicivita:(\boldr_{C\!I}\otimes \boldn_{\alpha e}) + m^{(0e)}_{\alpha}\boldn_{\alpha e}\right]
&
\boldm^{(0)} &= \boldf_m \left(\boldchi^{(-2)}\right)
&
\chi^{(-2)}_{\alpha} &= \frac{\boldn_{\alpha}}{l}\cdot\left[\boldtheta^{(-1J)} - \boldtheta^{(-1I)}\right]
\label{eq:LC2_O-1_b}
\end{align}
The degrees of freedom are particle rotations $\boldtheta^{(-1)}$, appearing also on the previous scale in equations~\eqref{eq:LC2_0-2}. Combining equations \eqref{eq:LC2_0-2} and \eqref{eq:LC2_O-1_b}, the elastic $y$-periodic solution becomes the following. The total rotations $\boldtheta^{(-1)}$ are zero and the displacements $\boldu^{(0)}$ are constant over $y$; consequently, the curvature and strain are zero, as well as the traction and couple traction $\bolde^{(-1)} = \boldchi^{(-2)} = \boldt^{(-1)} = \boldm^{(0)}=\bm{0}$. The volumetric strain $\varepsilon_V^{(-1)}$ from Eq.~\eqref{eq:LC2_0-2_kinematics} is then also zero using the same mathematical proof as presented in Eq.~\eqref{eq:zero_ev0}. Yet again, the validity of the elastic solution is extended also into the inelastic domain based on the same arguments as used in LC1.

The linear momentum balance is given by equation~\eqref{eq:LC1-t0} and the corresponding constitutive equation~\eqref{eq:const0_LC2} must be evaluated with strains on the higher scale. They can be written as
\begin{align}
\bm{0} &= \sum_{e\in I} A_e^{\star} t^{(0e)}_{\alpha}\boldn_{\alpha e} & \boldt^{(1)} &= \boldf_s \left(\bolde^{(1)},\,\varepsilon^{(1)}_V\right) \label{eq:LC2_O-1_c}
\end{align}
The kinematic equations~\eqref{eq:str0} and \eqref{eq:ev0X} are rewritten using total rotation~\eqref{eq:theta0}. The results of previously solved problems ($\boldtheta^{(-1)}=\bm{0}$ and $y$-constant $\boldu^{(0)}$ are substituted, as well as the identity $\x_{I\!J} = \boldr_{C\!I}-\boldr_{C\!J}$. 
\begin{subequations}
\begin{align}
e^{(0)}_{\alpha} &= 
\frac{\boldn_{\alpha}}{l}\cdot\left[\boldu^{(1)}_J - \boldu^{(1)}_I + \boldlevicivita: \left(\boldtheta^{(0)}_J \otimes \boldr_{C\!J} - \boldtheta^{(0)}_I\otimes \boldr_{C\!I} \right) + \x_{I\!J} \cdot \nabla_{X}\stackrel{\mathrm{s}}{\otimes}\boldu^{(0)} \right] 
\\
\varepsilon^{(0)}_V &= -\frac{1}{9  V_t} \left(\sum\limits_{I\in t}  A_I  \boldu^{(1)}_I \cdot \boldn_I\right) + \frac{1}{3} \mathrm{tr} \left( \nabla_{ X}  \cdot \boldu^{(0)} \right) 
\end{align}
\end{subequations}
In the previous equation, the symmetric gradient of displacements $\boldu^{(0)}$ appears from the difference between the full and antisymmetric gradients in Eq.~\eqref{eq:str0}.

Once again, the same situation as on the scale $\mathcal{O}(\eta^{-2})$ is encountered: the number of available equations is less than the unknown variables. 
The additional required equations appear on the next scale. 

At this point, it is convenient to rewrite the kinematic equations as standard discrete model equations with imposed eigen-strains.
\begin{subequations}
\label{eq:LC2_-1_c}
\begin{align}
e^{(0)}_{\alpha} &= 
\frac{\boldn_{\alpha}}{l}\cdot\left[\boldu^{(1)}_J - \boldu^{(1)}_I + \boldlevicivita: \left(\boldtheta^{(0)}_J \otimes \boldr_{C\!J} - \boldtheta^{(0)}_I\otimes \boldr_{C\!I} \right)\right] - \hat e_{\alpha} & \text{where}\quad \hat e_{\alpha} & = - \boldn_N \cdot \boldvarepsilon \cdot \boldn_{\alpha}
\\
\varepsilon^{(0)}_V &= -\frac{1}{9  V_t} \left(\sum\limits_{I\in t}  A_I  \boldu^{(1)}_I \cdot \boldn_I\right)  - \hat \varepsilon_V & \text{where}\quad \hat \varepsilon_V & = - \frac{1}{3} \mathrm{tr} \left( \boldvarepsilon \right) 
\end{align}
\end{subequations}
where $\boldvarepsilon=\nabla_{X}\stackrel{\mathrm{s}}{\otimes}\boldu^{(0)}$ is the symmetric displacement gradient, which serves as the strain tensor in Cauchy continuum. Clearly, we obtained part of the kinematic, constitutive, and balance equations of the standard discrete model Eqs.~(\ref{eq:disstraincurv},\ref{eq:volumetric_strain},\ref{eq:distconstitutive_eq},\ref{eq:LDPMbalance})

%%%%%%%%%%%%%%%%%%%%%%%%%%%%%%%%%%%%%%%%%%%%%%%%%%%%%%%%%%%%%%%%%%%%%%%%
\subsection{LC2: The \texorpdfstring{$\mathcal{O}(\eta^{0})$}{O(0)} problem}

Let us first write down the angular momentum balance equation~\eqref{eq:LC1-m0} along with the corresponding balance equation~\eqref{eq:const0_LC2} transferred to this scale since $\boldchi^{(-2)}=\bm{0}$. The kinematic equation \eqref{eq:curv-1} is rewritten with substituted Eq.~\eqref{eq:theta_expansion} and previously solved primary variables.
\begin{align} 
-V_I\boldz &=  \sum_{e\in I} A_e^{\star}\left[ t^{(0e)}_{\alpha}\boldlevicivita:(\boldr_{C\!I}\otimes \boldn_{\alpha e}) + m^{(1e)}_{\alpha}\boldn_{\alpha e}\right] & \boldm^{(1)} &= \boldf_m \left(\boldchi^{(-1)}\right) & 
\boldchi^{(-1)}_{\alpha} &= \frac{\boldn_{\alpha}}{l}\cdot\left( \boldtheta^{(0)}_J - \boldtheta^{(0)}_I\right)
\label{eq:LC2_O0_b}
\end{align}
Equations~\eqref{eq:LC2_-1_c} and \eqref{eq:LC2_O0_b}  correspond exactly to the kinematic (\ref{eq:disstraincurv} and \ref{eq:volumetric_strain}), constitutive~\eqref{eq:distconstitutive_eq}, and balance equations~\eqref{eq:LDPMbalance} of the standard discrete model with degrees of freedom $\boldu^{(1)}$ and $\boldtheta^{(0)}$. The only difference are eigen-strains  obtained as a~projection of symmetric displacement gradient and volumetric strain on the coarse scale. The linear momentum balance equation does not contain any loading term and is free of any inertia. The inertia component is also missing in the angular momentum balance, however, the volume couple $\boldz$ is present. 

The fine scale problem is therefore solved numerically as a~standard, \emph{steady-state} discrete model problem with degrees of freedom $\boldu^{(1)}$ and $\boldtheta^{(0)}$ loaded by projection of the Cauchy strain on coarse scale $\boldvarepsilon=\nabla_X \stackrel{\mathrm{s}}{\otimes} \boldu^{(0)}$, volumetric strain on coarse scale $\varepsilon_V = \text{tr}(\boldvarepsilon)/3$, and volume couple $\boldz$. The boundary conditions are given by the required periodicity of the degrees of freedom. An~additional boundary condition that prevents rigid body displacement is needed since the kinematic equations contain only differences of $\boldu^{(1)}$. The first non-zero displacement term is $\boldu^{(0)}$, all the higher displacements represent fluctuations with zero mean. Therefore, it should be imposed that the weighted average of the displacements over the entire fine scale be zero.
\begin{align}
\left\langle \boldu^{(1)} \right\rangle = \frac{1}{\VRVE} \sum_{I\in\VRVE} V_I \boldu^{(1)}_I = \bm{0}  \label{eq:LC2verage}
\end{align}
Similarly to LC1, this condition can be enforced as a~linear constraint, via the Lagrange multiplier, or advantageously during post-processing. The rigid-body rotation of the fine-scale problem is automatically prevented by periodicity, and also $\boldtheta^{(0)}$ is the highest non-zero term, so its mean is not required to vanish.

After describing the fine-scale problem that spans the scales $\mathcal{O}(\eta^{0})$ and $\mathcal{O}(\eta^{1})$, our focus shifts to the balance of linear momentum. Since the fine-scale linear momentum balance equation was already solved for $\boldu^{(1)}$, the next equation should determine the coarse-scale translation $\boldu^{(0)}$ considering the overall balance of the fine-scale model given by Eq.~\eqref{eq:LC1-t0}. The summed equation is identical to Eq.~\eqref{eq:LC1mac_linsummed} and reads
\begin{align}
\left\langle\rho\right\rangle \ddot{\boldu}^{(0)}  - \left\langle\boldb\right\rangle &= \frac{1}{\VRVE}\sum_{I\in\VRVE} \sum_{e\in I} A_e^{\star} t^{(1e)}_{\alpha} \boldn_{\alpha e}
\end{align}
The constitutive model is obtained by Taylor expansion and is identical to the left-hand side of Eq.~\eqref{eq:const_taylor_expansion}. The kinematics is also identical to LC1, the first and third relation from Eq.~\eqref{eq:strainsum} is used here. The very same approach as in LC1 is followed here as well, resulting in the definition of homogeneous continuous corse-scale balance equation and definition of stress tensor
\begin{align}
\left\langle\rho\right\rangle \ddot{\boldu}^{(0)}  - \left\langle\boldb\right\rangle &= \nabla_{X}\cdot \boldsigma & \text{where}\quad\boldsigma &= \frac{1}{\VRVE}\sum_{e\in\VRVE} l_e A_e^{\star} t^{(0e)}_{\alpha} \boldn_{N\!e} \otimes \boldn_{\alpha e} \label{eq:LC2_Cauchy_linear}
\end{align}
These equations are identical to the first parts of Eqs.~\eqref{eq:LC1_Cosserat_linear} and \eqref{eq:LC1_macrostress}. The coarse scale is clearly a~standard Cauchy continuum with degrees of freedom $\boldu^{(0)}$. The fine-scale model is the standard discrete model with degrees of freedom $\boldu^{(1)}$ and $\boldtheta^{(1)}$, and is loaded by projection of the symmetric strain tensor $\boldvarepsilon=\nabla_X \stackrel{\mathrm{s}}{\otimes} \boldu^{(0)}$, volumetric strain $\varepsilon_V=\mathrm{tr}(\boldvarepsilon)/3$, and volume couple $\boldz$. This result  corresponds to the findings presented in Ref.~\parencite{PradelThesis} for steady-state linear elasticity under hypothesis HS1. The inertia terms related to rotational degrees of freedom again do not appear on any relevant scale. The fine scale model is steady state; the coarse scale contains only inertia by translations $\boldu^{(0)}$.

\textcite{ForPra-01} pointed out that the stress tensor is not necessarily symmetric. To derive the same result here, one look again at the balance of angular momentum~\eqref{eq:LC1-m0}. However, this time the focus is on balance of the whole periodic fine-scale model. Following the very same path taken on scale $\mathcal{O}(0)$ in LC1, the summed momentum balance is written with respect to the origin of the reference system $y$.
\begin{align}
- \left\langle\boldz\right\rangle &= \frac{1}{\VRVE}\sum_{I\in\VRVE} \sum_{e\in I} A_e^{\star}\left[t^{(0e)}_{\alpha}\boldlevicivita:\left(\y_I+\boldr_{C\!I}\right)\otimes \boldn_{\alpha e} + m^{(1e)}_{\alpha}\boldn_{\alpha e}\right]
\end{align}
with constitutive and kinematic equations written in Eq.~\eqref{eq:LC2_O0_b} on the right-hand side. As in LC1, the vector $\y_I$ can be removed directly since $\sum_{e\in I} A_e^{\star} t^{(0e)}_{\alpha}\boldn_{\alpha e}=\bm{0}$ according to Eq.~\eqref{eq:LC2_O-1_c}. Transformation of double summation to single summation over all elements in the fine-scale model visit each element twice, each time with different orientation. Contrary to LC1, this time $^{I\!J}\boldm^{(1)}=^{I\!J}\boldm^{(1)}$ and the term with couple traction cancels out. The remaining terms are
\begin{align}
-\left\langle\boldz\right\rangle &= \boldlevicivita:\boldsigma
\label{eq:LC2_Cauchy_angular}
\end{align}
The stress tensor is symmetric only in the absence of an~external couple $\boldz$. Any volume couple load makes it non-symmetric. 

%%%%%%%%%%%%%%%%%%%%%%%%%%%%%%%%%%%%%%%%%%%%%%%%%%%%%%%%%%%%%%%%%%%%%%%%
\section{Heuristic case  HC3: combination of cases LC2 and LC1}
%%%%%%%%%%%%%%%%%%%%%%%%%%%%%%%%%%%%%%%%%%%%%%%%%%%%%%%%%%%%%%%%%%%%%%%%
The homogenization of a~discrete system with rotational degrees of freedom through an~asymptotic expansion results in two distinct cases depending on the scale at which the Cosserat length, $\ell_c$, lives. Limiting case LC2 ($\ell_c\sim d_c$) leads to the coarse-scale Cauchy continuum with a~standard discrete model on the fine scale, case  LC1 ($\ell_c\sim L_c$) provides the coarse-scale Cosserat continuum with decoupled discrete fine-scale model. Naturally, none of these asymptotic cases occurs in reality. \textcite{ForPra-01} developed a~general case, labeled HS3 by them and HC3 here, where the coarse-scale Cosserat continuum is linked to the fine-scale consisting of the standard discrete model. HC3 is a~heuristic approach, which does not have a~rigorous mathematical derivation at the moment. However, the authors speculate this heuristic solution to be the exact general solution of the homogenization problem valid for any value of the fine-scale Cosserat length scale. Incidentally, this solution coincides with the solutions presented in \cite{RezCus16} and \cite{EliCus22}, in which, however, some unjustified assumptions were made as discussed earlier in this paper. The possibility to derive exactly this general case is still an open topic of research.

The coarse scale is taken from the LC1 derivation, it is represented by balance equations~\eqref{eq:macrobalance_LC1} and the definition of coarse-scale flux variables~\eqref{eq:macroflux_LC1}. The degrees of freedom are $\boldu^{(0)}$ and $\boldtheta^{(0)}$, the kinematic equations are simply definitions of the Cosserat strain and curvature $\boldgamma$ and $\boldkappa$. 

The fine scale is taken from LC2; it is the standard discrete model without inertia effects with translational degrees of freedom $\boldu^{(1)}$. The fine-scale rotational degrees of freedom are denoted $\boldtheta^{(0)}$ in LC2, but this symbol is already used at coarse scale in LC1, therefore we will refer to it as $\boldtheta^{(0\star)}$. This inconsistency reflects the heuristic combination of the two results. The load is provided by both the eigen-strain, eigen-curvature, and volumetric eigen-strain given as projections of coarse-scale strain, curvature, and volumetric strain according to Eq.~\eqref{eq:LC1_eigen}. Periodic boundary conditions are applied. Furthermore, since both $\boldu^{(0)}$ and $\boldtheta^{(0\star)}$ are fluctuation fields, both should be zero on average according to Eq.~\eqref{eq:LC1_bc}. The first zero average is easy to satisfy during post-processing, it does not affect the solution, only restricts the rigid body translation of the fine-scale model. However, the constraint of zero mean rotations has to be applied through the Lagrange multiplier or a~linear constraint, as it now changes the fine-scale solution.

%%%%%%%%%%%%%%%%%%%%%%%%%%%%%%%%%%%%%%%%%%%%%%%%%%%%%%%%%%%%%%%%%%%%%%%%
\section{Verification using elastic two-dimensional models}
%%%%%%%%%%%%%%%%%%%%%%%%%%%%%%%%%%%%%%%%%%%%%%%%%%%%%%%%%%%%%%%%%%%%%%%%

To verify the derived equations, a~simple elastic discrete model in two dimensions is used. Since only elastic behavior is studied, there is no effect of volumetric strain. The model is identical to the one presented in Ref.~\parencite{EliCus25}. Circular aggregates are randomly placed in the domain without overlap, the diameters of these aggregates span the interval from 10\,mm to 4\,mm and their distribution follows the Fuller curve. The power/Lauguerre tessellation then generates the polygonal shape of the rigid bodies and the contact facets between them. The kinematic~\eqref{eq:disstraincurv} and balance equations~\eqref{eq:LDPMbalance} from the beginning of the paper are used. The constitutive equation is taken from Ref.~\parencite{EliCus25}
\begin{align}
t_N & =E_0 e_N & t_M &= \alpha E_0 e_M & m & = \beta \frac{E_0 A^2}{12} \chi \label{eq:discrete_constitutive}
\end{align}
where $E_0$ is the normal elastic modulus, $\alpha$ is the ratio between tangential and normal stiffness, and $\beta$ sets the bending stiffness. Usually, the bending stiffness parameter $\beta$ is set to 0 in the literature. However, it is used here to control the effective Cosserat length of the discrete system. According to the appendix of Ref.~\parencite{EliCus25}, $\beta=1$ corresponds to the case where traction is integrated exactly over the facet, i.e., all the couple traction is only due to nonuniform distribution of traction over the contact. In this work, the elastic parameters are set to $E_0=60$\,GPa and $\alpha=0.25$, while $\beta$ is kept variable, spanning the interval from $10^{-4}$ to $10^3$.

\subsection{Macroscopic elastic parameters}
\begin{figure}[!tb]
\centering
\includegraphics[width=5cm]{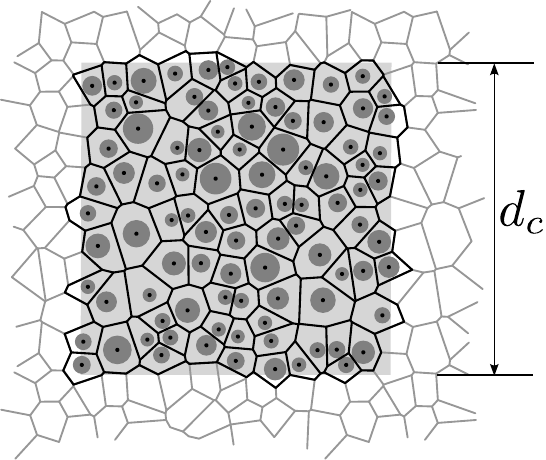}
\caption{Example of a~fine-scale model of size $d_c=100$\,mm with periodic internal structure.}
\label{fig:RVE}
\end{figure}

Square RVEs of three different sizes $d_c=50$\,mm, 100\,mm and 150\,mm are generated. They are generated fully periodically; that is, the placement of the circular aggregates and tessellation are implemented using periodic distance (see Refs.~\parencite{EliVor16,EliVor20,EliVor-20} and Fig.~\ref{fig:RVE}). The elastic RVEs are loaded by projections of the coarse-scale strain and curvature tensors $\boldvarepsilon$, $\boldgamma$  and $\boldkappa$, depending on the limiting case investigated. The loading proceed sequentially, setting each time one component equal to one while the others are zero. The resulting stress and couple stress components are collected and organized into a~second order tensor of elastic constants $D$. Since for each size there are 150 different period fine-scale geometries, there are in total 3$\times$150 such tensors evaluated. Since the coarse scale is assumed to be either Cauchy or Cosserat continuum, these tensors are matched by the corresponding tensors from continuum mechanics.
\begin{align}
\mathbf{D}_{\text{Cauchy}} &= \left(\begin{array}{ccc}
\lambda+2\mu & \lambda & 0\\
\lambda & \lambda+2\mu & 0\\
0 & 0 & \mu\end{array} \right) & \mathbf{D}_{\text{Cosserat}} &= \left(\begin{array}{cccccc}
\lambda+2\mu & \lambda & 0 & 0 & 0 & 0 \\
\lambda & \lambda+2\mu & 0 & 0 & 0 & 0 \\
0 & 0 & \mu+\mu_c & \mu-\mu_c & 0 & 0 \\
0 & 0 & \mu-\mu_c & \mu+\mu_c & 0 & 0 \\
0 & 0 & 0 & 0 & 4\mu\ell_c^2 & 0 \\
0 & 0 & 0 & 0 & 0 & 4\mu\ell_c^2
\end{array} \right) \label{eq:CosseratTensor}
\end{align}
where Lamé constants $\lambda$ and $\mu$ are related to the elastic modulus $E$ and Poisson's ratio $\nu$ through the expressions $E = \mu(3\lambda+2\mu)/\left(\lambda+\mu\right)$ and $\nu=\lambda/\left[2(\lambda+\mu)\right]$. The Cauchy tensor of elastic constants, $\mathbf{D}_{\text{Cauchy}}$, is the standard tensor for \emph{plane strain} elasticity, while the Cosserat tensor of elastic constants, $\mathbf{D}_{\text{Cosserat}}$, is taken from Ref.~\parencite{ZhaWan-05} and is also derived for \emph{plane strain} simplification. $\ell_c$ is the Cosserat length and $\mu_c$ is an additional parameter of the Cosserat material. Matching the computed tensors with the theoretical ones from Eq.~\eqref{eq:CosseratTensor} is done by means of mean square error minimization. 

\begin{figure}[!tb]
\centering
\includegraphics[width=\textwidth]{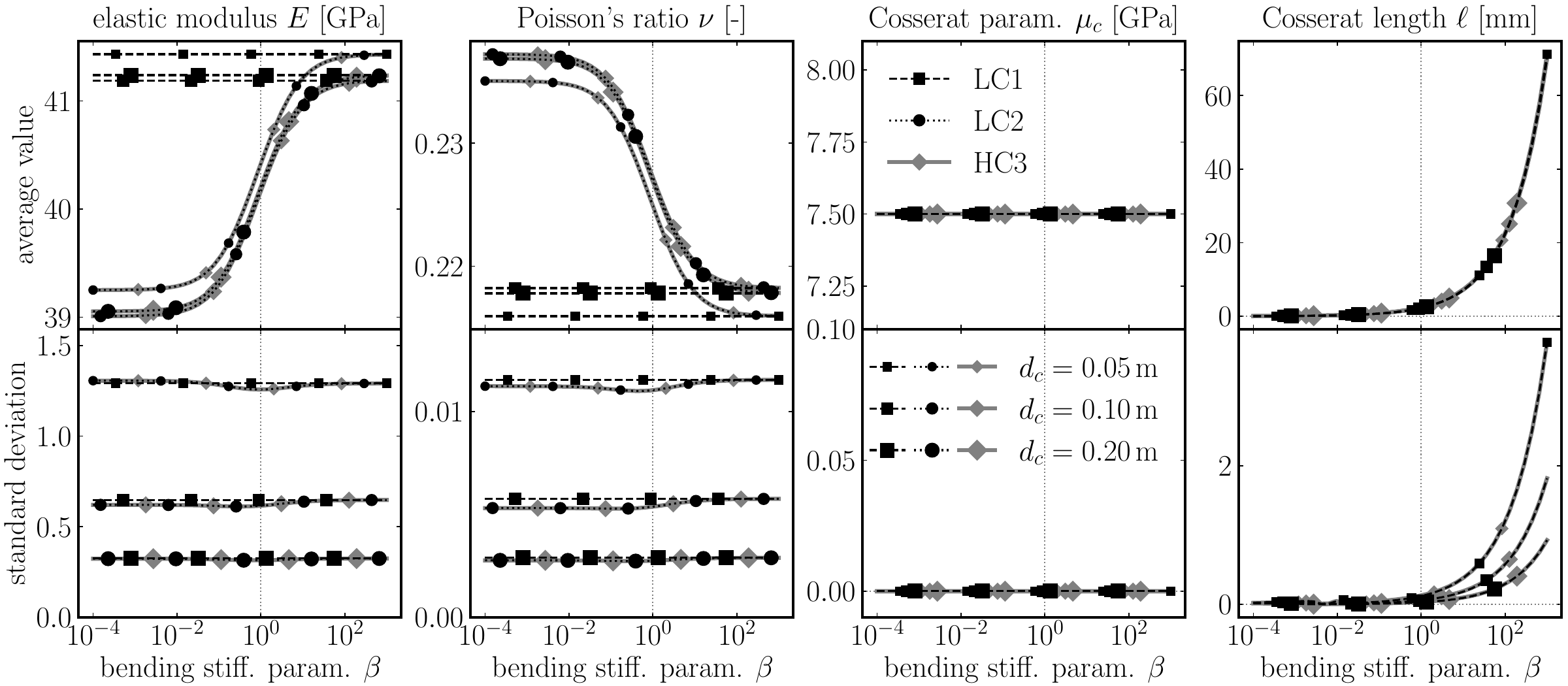}
\caption{Macroscopic elastic parameters computed from 150 fine-scale periodic models with different internal structure for each size. The upper row reports estimation of the mean value, the bottom row shows the standard deviation.}
\label{fig:RVE_statistics}
\end{figure}

Figure~\ref{fig:RVE_statistics} shows statistical results based on the 3$\times$150 sets of the computed elastic constants. In the upper row, the estimated mean values are plotted, the bottom row shows standard deviations, the horizontal axis always denotes the value of the bending stiffness parameter $\beta$. The extremely low values of $\beta$ result in systems with practically no bending stiffness, while the extremely high values of $\beta$ completely eliminate all curvatures at the facets.

The first important observation is that the mean value of all parameters is independent of the fine-scale size $d_c$. The small differences are caused by an~insufficient sample size and would diminish with considering more samples of fine-scale internal structure. As expected, the standard deviations decrease rapidly with increasing $d_c$. The Cosserat parameter $\mu_c$ exhibits a~very stable value, almost no variation even for the smallest fine-scale size $d_c=0.05$\,m.

Increasing $\beta$ in case LC2 gradually changes the coarse-scale elastic behavior of the system, the elastic modulus increases, and the Poisson's ratio decreases. Both of them converges to a~limit value as the rotations lock to a~constant value over the whole fine-scale model (not necessarily zero).   

For the limiting case LC1, there is no effect of $\beta$ on the elastic modulus and Poisson's ratio. These parameters come from the coarse-scale stress tensor of the decoupled fine-scale where rotations have no effect on tractions and moments of tractions are disregarded. One can see it as an~asymptotic behavior of LC2 with $\beta\rightarrow\infty$, that is, the rotations are locked. Indeed, the elastic modulus and Poisson's ratio of LC1 match the asymptotic values of LC2. Furthermore, case  LC1 shows a~substantial increase in coarse-scale Cosserat length, $\ell_c$. This length would grow further to infinity with increasing $\beta$.

Finally, the heuristic case HC3 can match the Cosserat parameters $\mu_c$ and $\ell_c$ provided by case LC1 and the Cauchy parameters $E$ and $\nu$ produced by case LC2. This correspondence is quite precise.

The authors do not know of any material for which the value $\beta$ should approach the large values included in the study. The realistic value might be around $\beta=1$ where the bending stiffness is purely caused by the traction acting on eccentricity. This case is emphasized in Fig.~\ref{fig:RVE_statistics} by a~vertical dotted line and clearly lies in the transitional region. Setting $\beta=0$ as is done in the majority of research papers dealing with discrete models underestimates slightly the coarse-scale elastic modulus of the system and overestimates its Poisson's ratio.  However, the Cosserat length for $\beta=1$ is low and the case LC2 with the Cauchy continuum at the coarse scale is clearly a~better choice than case LC1 assuming Cosserat effects.

\begin{figure}[!tb]
\centering
\includegraphics[width=11cm]{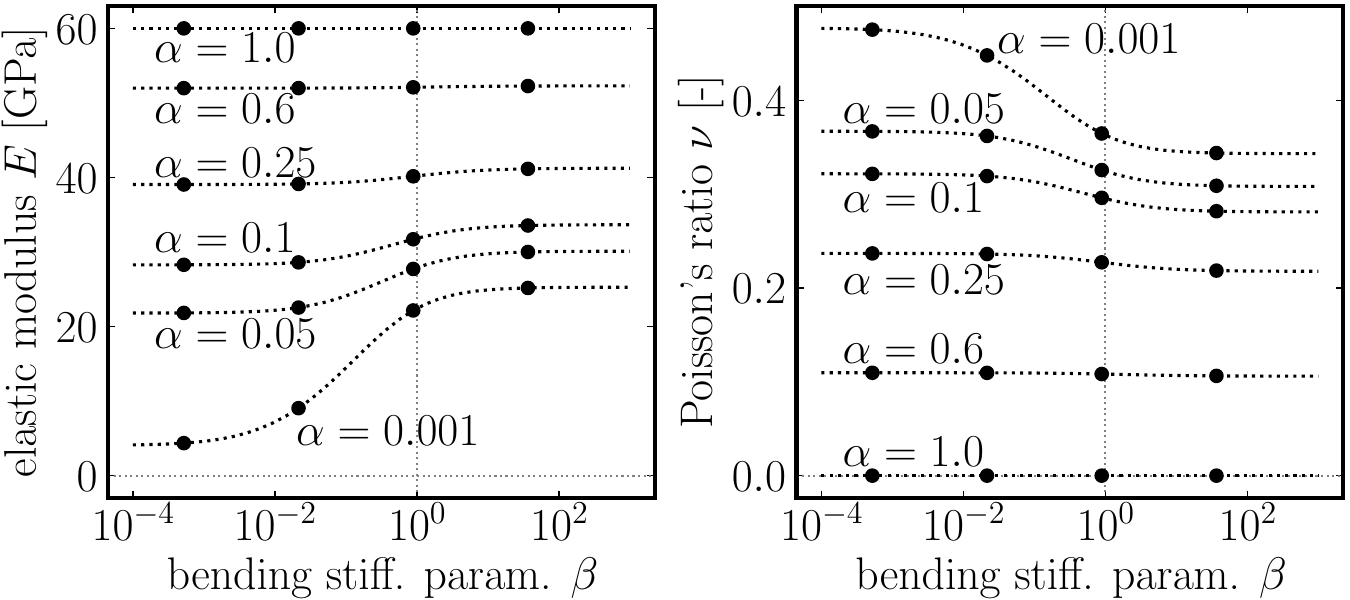}
\caption{Mean value estimates of coarse-scale elastic parameters computed from 150 fine-scale periodic models with different internal structure. Only size $d_c=0.2$\,m and limiting case LC2 is shown.}
\label{fig:RVE_alpha}
\end{figure}

The relative difference between the upper and lower LC2 asymptotic limits ($\beta\rightarrow 0$ and $\beta\rightarrow\infty$) of the elastic modulus and Poisson's ratio is only about 5.6\,\% and 8.7\,\%, respectively. However, this difference might increase or decrease significantly depending on the parameter $\alpha$, the ratio between the tangential and normal stiffness of the contacts. The simple study in Fig.~\ref{fig:RVE_alpha} changes the parameter $\alpha$ in the homogenization scheme LC2 and computes the averages of the coarse-scale elastic modulus and Poisson's. The maximum difference is achieved for $\alpha=10^{-4}$, the elastic modulus differs by 513\,\% and the Poisson's ratio by 39\,\%. These values can be even higher as $\alpha$ decreases further to 0. The unstable system with $\alpha=\beta=0$ exhibits zero coarse-scale elastic modulus and Poisson's ratio is 0.5, the asymptotic behavior of the other extreme with $\alpha=0$ and $\beta=\infty$ is not known, but it is clear that the differences become maximized.  On the other hand, setting $\alpha=1$ eliminates all rotations in the system \parencite{Eli17,Eli20} and therefore the value $\beta$ becomes irrelevant. 

\subsection{Steady-state discrete system}
The next verification compares the \emph{full} discrete system with homogenized models featuring RVEs acting according to cases LC2, LC1, and HC3. The problem at hand is a~two dimensional cantilever of depth $D=1$\,m and length 6\,m with restricted horizontal and vertical displacements, as well as rotations on the left-hand side. On the right-hand side, the boundary behaves as a~single rigid body interconnecting all the involved translational and rotational degrees of freedom. The loading force acts on this rigid body in the vertical direction, where the vertical displacement is also measured. 

The \emph{full} model is created by random placement of non-overleaping circular particles with Fuller sieve curve and diameters between 10 and 4 millimeters. The rigid bodies and contacts between them are then given by the power tessellation. In total, the \emph{full} model contains approx.~172\,000 degrees of freedom. The elastic material parameters are $E_0=60$\,GPa, $\alpha=0.25$, and $\beta$ vary. Six such models are created with different internal structures, and the structural bending stiffness evaluated as the loading force divided by the deflection is averaged from these samples. 

The homogenized stiffness according to the three derived schemes LC2, LC1, and HC3 is calculated next. On the coarse scale, square bilinear finite elements are used. For limiting case LC2, these are standard isoparametric Cauchy quadrilaterals. In cases LC1 and HC3, isoparametric bilinear Cosserat elements according to \textcite{ZhaWan-05} are employed. Three variants of square finite element sizes are considered: $D/20$, $D/10$, and $D/5$, the mean stiffness is computed as an~average over 150 evaluations using 150 RVEs with different internal structure for each element size. This time, only $d_c=0.2$\,m is reported as the other exhibit identical results only less accurate. 

\begin{figure}[!tb]
\centering
\includegraphics[width=\textwidth]{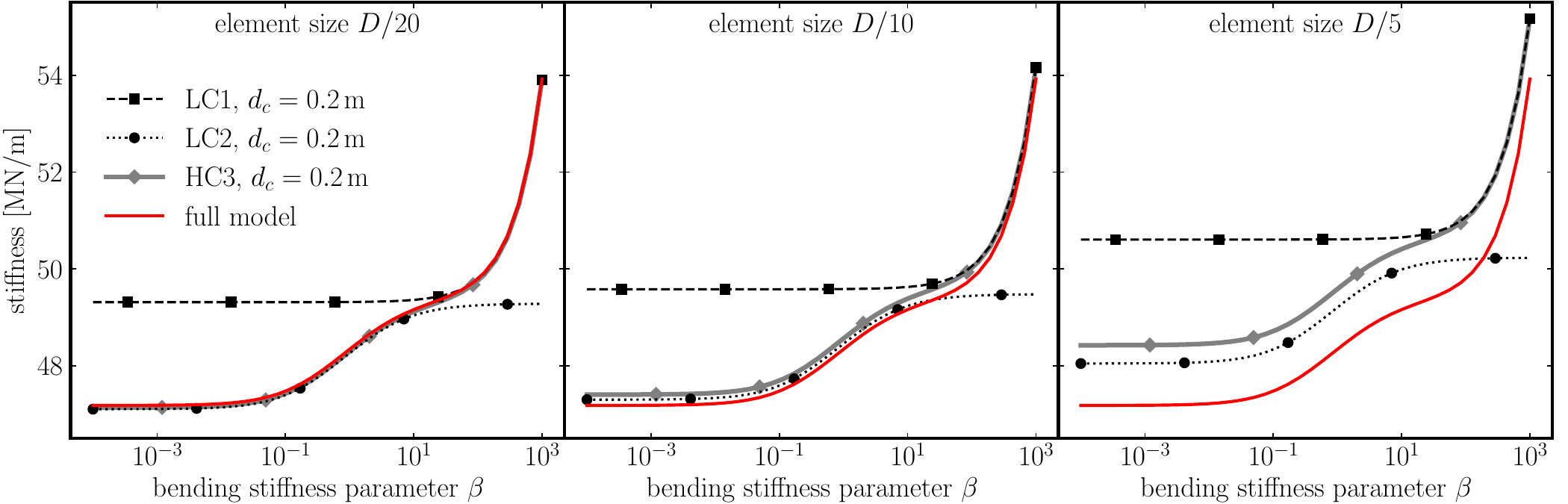}
\caption{Structural bending stiffness of the \emph{full} discrete model and \emph{homogenized} models with different homogenization schemes. The averages are taken over 6 samples (\emph{full} model) or 150 samples (\emph{homogenized} model), respectively.}
\label{fig:stiffness}
\end{figure}

The results are plotted in Fig.~\ref{fig:stiffness}. The best results are obtained for the smallest finite elements for which the coarse-scale kinematics is the least restricted. For large $\beta$ values (from approx. 100), case LC1 is in good agreement with the \emph{full} model. For low values of $\beta$ (up to approximately 10), the correct limiting case becomes LC1. The heuristic case HC3 seems to match the \emph{full} model for all $\beta$ variants. Also, note that the asymptotic limit $\beta\rightarrow 0$ of LC1 is equal to the asymptotic limit $\beta\rightarrow \infty$ of LC2. This was already seen in the results in Fig.~\ref{fig:RVE_statistics}, in both cases the rotations are not involved in the evaluation of the traction. LC1 decouples them directly at the fine scale, LC2 eliminates them by setting $\beta$ to some high value.

When using larger finite elements, the \emph{homogenized} models become stiffer because their coarse-scale kinematics is more restricted. Interestingly, as the element size increases, HC3 becomes significantly stiffer than LC2 for low $\beta$ values, and the asymptotic limits of LC2 and HC3 no longer correspond. The Cauchy continuum in the case LC2 uses a~symmetric part of the strain tensor and is therefore kinematically less restricted compared to the Cosserat continuum, which distinguishes between shear strains $\varepsilon_{ij}$ and $\varepsilon_{ji}$. This restriction becomes more pronounced for larger elements; the coarse-scale Cosserat system does not have enough degrees of freedom to compensate for the restricted independent rotations.

\subsection{Transient discrete system}

The cantilever in the previous example is now loaded by force $100$\,kN imposed suddenly at time 0\,s. The transient solution is computed with the help of the generalized-$\alpha$ method of \textcite{ChuHul93}. The time step is set to 1\,ms, and the spectral radius 0.8 is considered. The \emph{full} model uses the consistent mass matrix that includes all inertia terms from Eq.~\eqref{eq:LDPMbalance}, density $\rho$ is 2400\,kg/m$^3$. All \emph{homogenized} models have steady-state fine-scale behavior, their only inertia component is in the coarse-scale linear momentum balance equations~\eqref{eq:LC2_Cauchy_linear} and \eqref{eq:LC1_Cosserat_linear}. 

\begin{figure}[!tb]
\centering
\includegraphics[width=14cm]{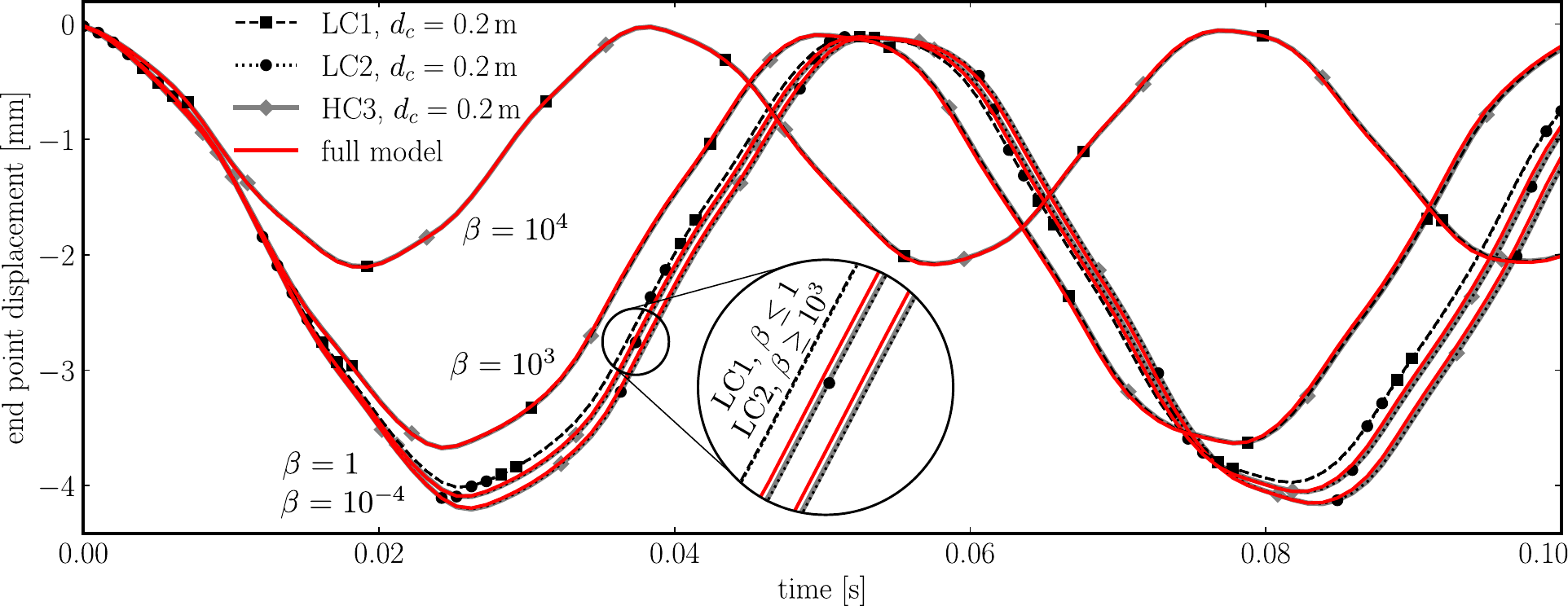}
\caption{Structural bending stiffness of the \emph{full} discrete model and \emph{homogenized} models with different homogenization schemes. The averages are taken over 6 samples (\emph{full} model) or 150 samples (\emph{homogenized} model), respectively.}
\label{fig:dynamics}
\end{figure}

Figure~\ref{fig:dynamics} plots the evolution of the vertical displacement of the cantilever end point in time, four different cases with $\beta=10^{-4}$, 1, $10^3$, and $10^4$ are shown. The \emph{full} model becomes stiffer, and therefore the amplitude decreases and the frequency increases with increasing $\beta$. The homogenized case HC3 matches well all the \emph{full} models. The case LC2 with the Cauchy coarse scale can reproduce low $\beta$ values, and the case LC1 with the Cosserat coarse scale provides an~excellent approximation for large $\beta$ values. The figure shows that the response of the LC2 scheme for high values of $\beta$ is equal to the response of the LC1 scheme for low $\beta$ values. This has already been observed in the previous verification examples; it is the response of a~mechanical discrete system in which the \emph{independent} particle rotations are either irrelevant (LC1) or restricted (LC2). 

\section{Verification using inelastic three-dimensional model}

Until now, all simulations were performed with an~elastic two-dimensional model. To verify the derived homogenization also for inelastic constitutive functions, a~three-dimensional model of the dogbone specimen is eccentrically loaded in tension under steady-state assumption. Creation of the internal geometry of the specimen is identical to the previous two-dimensional cases (sequential placement of non-overlapping particles generated according to the Fuller curve and power/Laguerre tessellation). The rotational stiffness is now completely omitted, i.e., parameter $\beta=0$. Consequently, only the limiting case LC2 resulting in the Cauchy continuum is used for homogenization. In the homogenized version, standard tri-linear isoparametric brick elements with 8 integration points are used.

\begin{figure}[!tb]
\centering
\includegraphics[width=14cm]{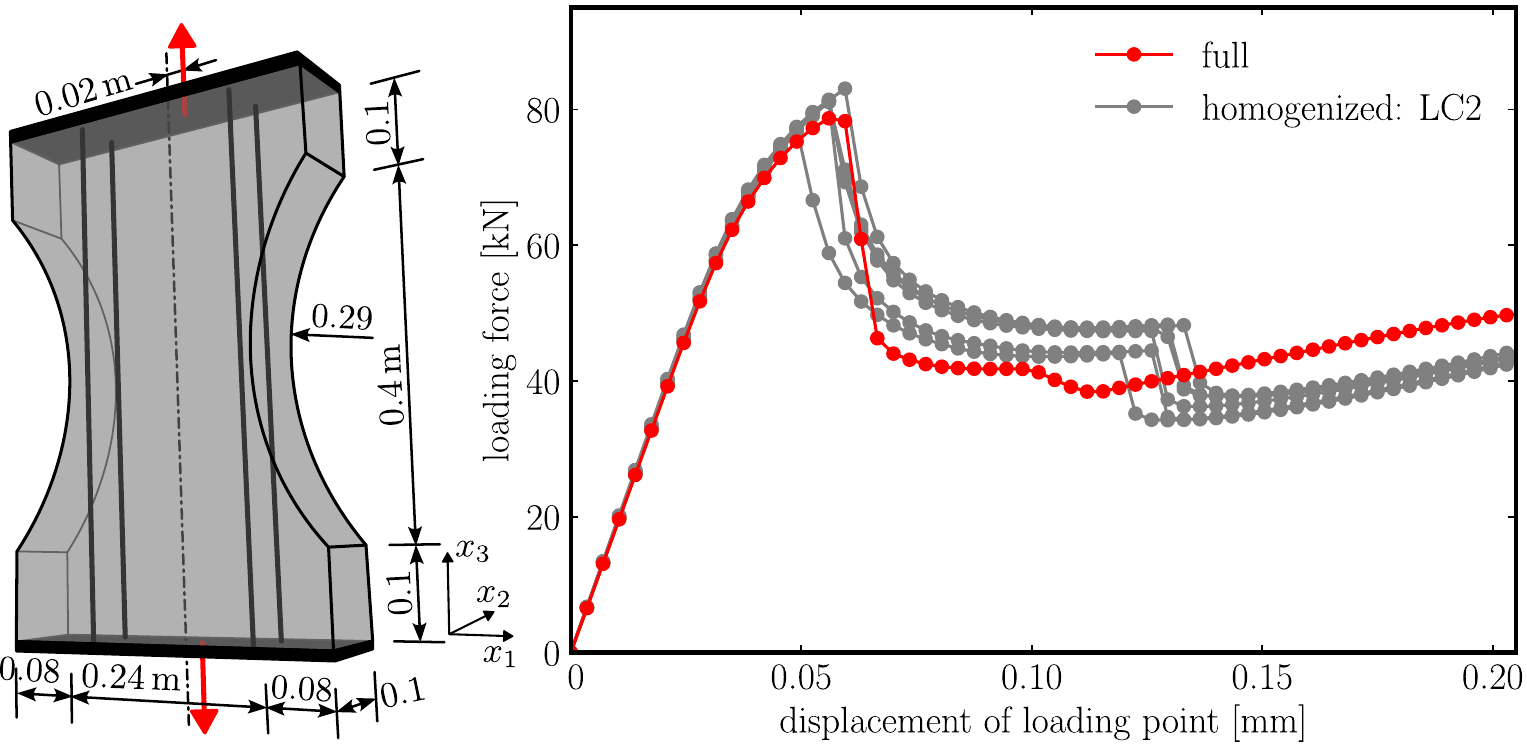}
\caption{Dimensions of the dogbone specimen loaded in eccentric tension and comparison of global responses of the \emph{full} and \emph{homogenized} models.}
\label{fig:dogbone_response}
\end{figure}

The dimensions of the dogbone specimen are specified in Fig.~\ref{fig:dogbone_response}. There are two rigid platens attached to the top and bottom faces. The translations of the rigid platens are restricted in both horizontal directions ($x_1$ and $x_2$) and their rotations around the $x_1$ and $x_3$ axes are restricted as well. Rotations around $x_2$ are free and translations in the $x_3$ direction are prescribed, so the specimen is loaded in eccentric tension.

There are also 4 steel reinforcing bars placed symmetrically around the central axis, the mutual distance of them is 0.204\,m in $x_1$ direction and 0.064\,m in $x_2$ direction.  The diameter of each steal rebar is 6\,mm. Their are implemented as elastic Timoshenko beams with Young's modulus 210\,GPa. In the \emph{full} model, these reinforcing bars are connected by a rigid arm constraint to all the particles that lie in their path~\parencite{BolEli-21}. For the homogenized model there is one hanging node constrained by shape functions at each finite element crossing the reinforcement.

\begin{figure}[!tb]
\centering
\includegraphics[width=\textwidth]{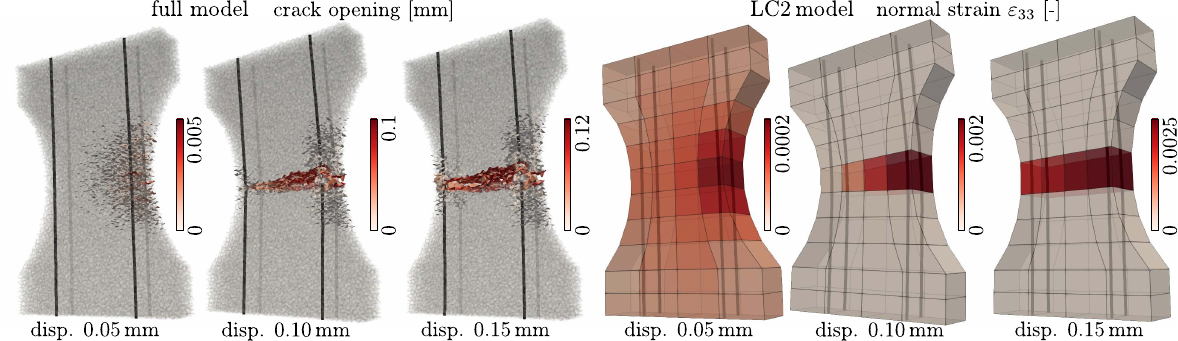}
\caption{Cracks developed in the \emph{full} and \emph{homogenized} models at loading displacements 0.05, 0.10, and 0.15\,mm.}
\label{fig:dogbones}
\end{figure}

The constitutive model of concrete in the elastic regime is identical to the previous two-dimensional model (Eq.~\ref{eq:discrete_constitutive}), $E_0=60$\,GPa, $\alpha=0.25$, $\beta=0$. Inelastic behavior is based on damage mechanics, each component of traction is multiplied by a factor $1-d$ where $d$ stands for a~non-decreasing damage parameter between 0 (intact) and 1 (completely disintegrated). Equation~\ref{eq:discrete_constitutive} then transforms into
\begin{align}
t_N & =(1-d)E_0 e_N & t_M &= (1-d)\alpha E_0 e_M & t_L &= (1-d)\alpha E_0 e_L 
\end{align}
The calculation of the damage parameter is not described here. The interested reader is referred to publications~\parencite{Eli16,EliCus22}, where the exactly identical model is used and is thoroughly described. It has only two additional parameters (above the elastic parameters), namely tensile strength and mesoscale fracture energy in tension. These parameters are arbitrarily chosen to be 3\,MPa and 50\,J/m$^2$.

It is well known that strain localization is a phenomenon that cannot be homogenized \textcite{GitAsk-07}. Instead of implementing some advanced technique available for example in Refs.~\parencite{Ung13,OliCai-15,TurHoo-18,TanDon-19}, much simpler remedy is used. As shown in \textcite{RezCus16,RezZho-17,EliCus22}, the fracture can be included if the size associated with the integration point in the coarse model in the direction perpendicular to the crack is the same as size of the fine scale model. The periodic fine scale models in three dimensions have size $0.03\times0.3\times0.3$\,m$^3$, therefore the size of finite elements is set to 0.06\,m along the $x_3$ direction.

Since the fine-scale model is relatively small, six of these models are generated differing by the internal position of particles and the coarse-scale simulation is repeated 6 times. Global response of these homogenized models are shown in Fig.~\ref{fig:dogbone_response} along with on simulation with the full discrete model. Developed cracks are compared in Fig.~\ref{fig:dogbones}. Both of these comparisons show reasonable correspondence and therefore verify the LC2 homogenization for inelastic constitutive functions.

\section{Conclusions}

This study applied asymptotic expansion homogenization to discrete models with rotational degrees of freedom. Following the work of \parencite{ForPra-01} on the homogenization of heterogeneous Cosserat continua, a~mathematically sound derivation is presented that leads to three different homogenization schemes.  The resulting equations are identical those derived in~\textcite{PradelThesis} for discrete systems. 
\begin{description}

\item[Limiting case LC1] For large bending stiffness $\beta$, the effective Cosserat length becomes comparable to the structural size, and the Cosserat continuum appears at the coarse scale. The fine scale emerges decoupled because the local traction moments are negligible compared to the couples. The tractions in the linear momentum balances are computed on a~discrete model with zero rotations, while the couples in angular momentum balances omit the effect of tractions. Projections of both the Cosserat strain and curvature tensors as well as the volumetric strain scalar provide the fine-scale loads. 

\item[Limiting case LC2] Assuming that the bending stiffness parameter $\beta$ is small enough so that the effective Cosserat length is comparable to the size of RVE, the discrete models homogenize to the Cauchy continuum. The coarse-scale stress tensor is symmetric unless some external volume couple load is imposed. The fine scale becomes the standard discrete model with periodic boundary conditions loaded by projection of the coarse-scale, symmetric strain tensor, and volumetric strain scalar.

\item[Heuristic case HC3] One can combine the results of LC2 and LC1 into a~heuristic solution where the coarse-scale is taken from the case LC2 and the fine scale from case LC1. However, one needs to impose both translations and rotations to be zero on average over the fine-scale model. The volume average of translations can be updated during post-processing, but the conditions for rotations must be imposed directly in the solver. It is worth nothing that HC3 corresponds exactly to the homogenization scheme presented in \cite{RezCus16} in which, however, the derivation was based on a number of assumptions on the rigid rotation of the RVE that were inconsistent with the periodicity of problem. Nevertheless, the authors speculate that HC3 is indeed the exact, general solution even though a formal proof does not exist yet.
\end{description}

In addition to these  previously published results, the paper also delivers the following novel observations. 
\begin{itemize}
\item Asymptotic homogenization is performed on a~transient system with all relevant inertial terms. It is shown that all three schemes result in steady-state fine scale, and the coarse scale keeps the inertia only with respect to translations.
\item In contrast to~\textcite{ForPra-01,PradelThesis}, the rotations are divided into the \emph{independent} part, for which the asymptotic expansion is performed, and the \emph{dependent} part, which are asymptotically expanded only through translations. Such treatment directly leads to symmetric and antisymmetric gradients at the coarse scale and also reveals the internal composition of the rotational degrees of freedom at different scales. The symmetric gradients in Cauchy continuum directly follows from this decomposition.
\item  The homogenization is derived for an~arbitrary constitutive function. An~effect of volumetric strain on tractions is also considered. Homogenization for inelastic material is verified by simulation crack initiation and propagation in a~reinforced concrete specimen.
\item The numerical results in elastic regime show that the HC3 scheme closely matches the \emph{full} model for any value of the bending stiffness parameter $\beta$. The other two mathematically rigorous schemes, LC2 and LC1, provide good agreement only in their respective domains. These are approximately $\beta<10$ for LC2 and $\beta>100$ for LC1. The authors assume that virtually all of the real materials shall have $\beta$ close to one, the case LC2 (coarse-scale Cauchy continuum) can be used whenever realistic numerical simulations are conducted.
\item The asymptotic expansion homogenization looks at the system with \emph{asymptotically small} RVEs. The expression of the coarse-scale couple stress~\eqref{eq:LC1_macrocouplestress} therefore does not contain any effect of traction on (infinitely small) eccentricity. Such an~effect would appear only when considering \emph{finite size} fine-scale problem, as in Refs.~\parencite{ChaLia-90} and \parencite[page 165]{Vard18_book} or in the recent paper~\parencite{EliCus25} of the authors.
\end{itemize}
The results presented in previous papers by the authors regarding the  mechanical discrete model homogenization~\parencite{RezCus16,RezZho-17,RezAln-19,EliCus22} are corrected now. However, their conclusions remain valid in the realistic setting where $\beta$ is low and provide results indistinguishable from the present theory. The Cosserat effects derived in those papers becomes negligible then and homogenization scheme LC2 is recovered. 

All simulations conducted in this article were performed using open source software named \emph{Open Academic Solver} (OAS)\footnote{\url{https://gitlab.com/kelidas/OAS}} developed at Brno University of Technology.

\appendix
\section{Asymptotic expansions of strain, volumetric strain, and curvature \label{sec:strain_expansion}}

Starting from Eqs.~\eqref{eq:disstraincurv}, replacing $\boldtheta$ by Eq.~\eqref{eq:rot_decomp}, using the asymptotic expansions of the displacements~\eqref{eq:expansion_u} and independent rotation~\eqref{eq:expansion_phi} as well as the Taylor expansion~\eqref{eq:Taylorseries}, and substituting Eq.~\eqref{eq:expansion_omega} contracted with Levi-Civita tensor according to Eq.~\eqref{eq:antisymmetric}, one arrives at equations~\eqref{eq:strain_expansion} with individual components reading
\begin{subequations} \label{eq:gechi_expansion}
\begin{align}
\tilde e^{(-1)}_{\alpha} &=
\frac{n^{\alpha}_i}{\tilde{l}}\left[\tilde u^{(0J)}_i - \tilde u^{(0I)}_i +\left(\levicivita_{ijk} \tilde \varphi^{(-1J)}_j - \nabla^a_{\tilde y_i}\tilde u^{(0J)}_k\right) \tilde{r}^{C\!J}_k - \left(\levicivita_{ijk} \tilde \varphi^{(-1I)}_j - \nabla^a_{\tilde y_i}\tilde u^{(0I)}_k\right)  \tilde{r}^{C\!I}_k 	\right]  \label{eq:str-1}  
\\
\tilde e^{(0)}_{\alpha} &= 
\frac{n^{\alpha}_i}{\tilde{l}}\left[\tilde u^{(1J)}_i - \tilde u^{(1I)}_i + \tilde x^{I\!J}_j \nabla_{\tilde X_j} \tilde u^{(0J)}_i  + \left(\levicivita_{ijk} \left(\tilde \varphi^{(0J)}_j +   \tilde x^{I\!J}_l \nabla_{\tilde X_l} \tilde \varphi^{(-1J)}_j\right) - \nabla^a_{\tilde X_i}\tilde u^{(0J)}_k - \nabla^a_{\tilde y_i}\left(\tilde u^{(1J)}_k + \tilde x^{I\!J}_l\nabla_{\tilde X_l}\tilde u^{(0J)}_k   \right)\right) \tilde{r}^{C\!J}_k \right. \nonumber \\ &\left. - \left(\levicivita_{ijk}\tilde\varphi^{(0I)}_j - \nabla^a_{\tilde X_i}\tilde u^{(0I)}_k - \nabla^a_{\tilde y_i}\tilde u^{(1I)}_k\right)\tilde{r}^{C\!I}_k \right] \label{eq:str0} 
\\
\tilde e^{(1)}_{\alpha} &= 
\frac{n^{\alpha}_i}{\tilde{l}}\left[ \tilde u^{(2J)}_i - \tilde u^{(2I)}_i + \tilde x^{I\!J}_j\nabla_{\tilde X_j} \tilde u^{(1J)}_i + \frac{1}{2}  \tilde x^{I\!J}_j \tilde x^{I\!J}_k \nabla_{\tilde X_j} \nabla_{\tilde X_k} \tilde u^{(0J)}_i +	
\left(\levicivita_{ijk} \left(\tilde \varphi^{(1J)}_j +   \tilde x^{I\!J}_l \nabla_{\tilde X_l} \tilde \varphi^{(0J)}_j +  \frac{1}{2} \tilde x^{I\!J}_l \tilde x^{I\!J}_m \nabla_{\tilde X_l} \nabla_{\tilde X_m} \tilde \varphi^{(-1J)}_j\right) \right.\right. \nonumber \\& \left.\left. - \nabla^a_{\tilde X_i}\left(\tilde u^{(1J)}_k + \tilde x^{I\!J}_l\nabla_{\tilde X_l}\tilde u^{(0J)}_k \right) - \nabla^a_{\tilde y_i}\left(\tilde u^{(2J)}_k +  \tilde x^{I\!J}_l\nabla_{\tilde X_l}\tilde u^{(1J)}_k + \frac{1}{2} \tilde x^{I\!J}_l \tilde x^{I\!J}_m\nabla_{\tilde X_l}\nabla_{\tilde X_m}\tilde u^{(0J)}_k\right) \right) \tilde{r}^{C\!J}_k \right.\nonumber\\ &\left.- \left(\levicivita_{ijk}\tilde\varphi^{(1I)}_j - \nabla^a_{\tilde X_i}\tilde u^{(1I)}_k - \nabla^a_{\tilde y_i}\tilde u^{(2I)}_k\right)\tilde{r}^{C\!I}_k\right] \label{eq:str1} 
\\
\tilde\chi^{(-2)}_{\alpha} &=\frac{n^{\alpha}_i}{\tilde{l}}\left[\tilde\varphi^{(-1J)}_i - \tilde\varphi^{(-1I)}_i + \frac{1}{2}\levicivita_{ijk}\nabla_{\tilde y_j}\left(\tilde u^{(0J)}_k -\tilde u^{(0I)}_k \right)
\right] \label{eq:curv-2} 
\\
\tilde\chi^{(-1)}_{\alpha} &= \frac{n^{\alpha}_i}{\tilde{l}}\left[ \tilde\varphi^{(0J)}_i - \tilde\varphi^{(0I)}_i +  \tilde x^{I\!J}_j \nabla_{\tilde X_j}\tilde\varphi^{(-1J)}_i + \frac{1}{2}\levicivita_{ijk}\left(\nabla_{\tilde X_j}\left(\tilde u^{(0J)}_k -\tilde u^{(0I)}_k \right) + \nabla_{\tilde y_j}\left(\tilde u^{(1J)}_k + \tilde x^{I\!J}_l \nabla_{\tilde X_l} \tilde u^{(0J)}_k-\tilde u^{(1I)}_k \right) \right) \right] \label{eq:curv-1} 
\\
\tilde \chi^{(0)}_{\alpha} &= \frac{n^{\alpha}_i}{\tilde{l}}\left[\tilde \varphi^{(1J)}_i - \tilde \varphi^{(1I)}_i +
\tilde x^{I\!J}_j\nabla_{ \tilde X_j} \tilde\varphi^{(0J)}_i  + \frac{1}{2} \tilde x^{I\!J}_j \tilde x^{I\!J}_k \nabla_{\tilde X_j} \nabla_{\tilde X_k} \tilde \varphi^{(-1J)}_i  +  \frac{1}{2}\levicivita_{ijk}\left(\nabla_{\tilde X_j}\left(\tilde u^{(1J)}_k + \tilde x^{I\!J}_l \nabla_{\tilde X_l} \tilde u^{(0J)}_k-\tilde u^{(1I)}_k \right) \right.\right. \nonumber \\ & \left.\left. + \nabla_{\tilde y_j}\left(\tilde u^{(2J)}_k +  \tilde x^{I\!J}_l\nabla_{\tilde X_l}\tilde u^{(1J)}_k + \frac{1}{2} \tilde x^{I\!J}_l \tilde x^{I\!J}_m\nabla_{\tilde X_l}\nabla_{\tilde X_m}\tilde u^{(0J)}_k -\tilde u^{(2I)}_k \right) \right) \right] \label{eq:curv0}  
\\
\tilde \chi^{(1)}_{\alpha} &= \frac{n^{\alpha}_i}{\tilde l}\left[\tilde \varphi^{(2J)}_i - \tilde \varphi^{(2I)}_i +
\tilde x^{I\!J}_j\nabla_{ \tilde X_j} \tilde\varphi^{(1J)}_i  + \frac{1}{2} \tilde x^{I\!J}_j \tilde x^{I\!J}_k \nabla_{\tilde X_j} \nabla_{\tilde X_k} \tilde \varphi^{(0J)}_i  + \frac{1}{6} \tilde x^{I\!J}_j \tilde x^{I\!J}_k  \tilde x^{I\!J}_l \nabla_{\tilde X_j} \nabla_{\tilde X_k} \nabla_{\tilde X_l} \tilde \varphi^{(-1J)}_i \right. \nonumber \\ &\left. +  \frac{1}{2}\levicivita_{ijk}\left(\nabla_{\tilde X_j}\left(\tilde u^{(2J)}_k + \tilde x^{I\!J}_l \nabla_{\tilde X_l} \tilde u^{(1J)}_k + \frac{1}{2} \tilde x^{I\!J}_l \tilde x^{I\!J}_m \nabla_{\tilde X_l} \nabla_{\tilde X_m} \tilde u^{(0J)}_k -\tilde u^{(2I)}_k \right) \right.\right. \nonumber \\ & \left.\left. + \nabla_{\tilde y_j}\left(\tilde u^{(3J)}_k +  \tilde x^{I\!J}_l\nabla_{\tilde X_l}\tilde u^{(2J)}_k + \frac{1}{2} \tilde x^{I\!J}_l \tilde x^{I\!J}_m\nabla_{\tilde X_l}\nabla_{\tilde X_m}\tilde u^{(1J)}_k + \frac{1}{6} \tilde x^{I\!J}_l \tilde x^{I\!J}_m \tilde x^{I\!J}_n \nabla_{\tilde X_l}\nabla_{\tilde X_m}\nabla_{\tilde X_n}\tilde u^{(1J)}_k -\tilde u^{(3I)}_k \right) \right) \right] \label{eq:curv1} 
\end{align}
\end{subequations}
Components of the antisymmetric part of the gradient, $(\nabla\stackrel{\mathrm{a}}\otimes\bullet)_{ij}$, are denoted $\nabla^a_{X_i}\bullet_j$.

The volumetric strain components are derived from Eq.~\eqref{eq:volumetric_strain} applied to the tetrahedron $t$ of volume $V_t$ with vertices $I$, $J$, $K$, and $L$, using expansions~\eqref{eq:expansion_u} and \eqref{eq:Taylorseries_u}. The individual components are
\begin{subequations}
\label{eq:ev_expansion}
\begin{align}
\tilde\varepsilon^{(-1)}_V &= -\frac{1}{9\tilde V_t} \sum\limits_{I\in t} \tilde A_I \tilde u^{(0I)}_i n^I_i \label{eq:ev-1}  \\
\tilde\varepsilon^{(0)}_V &= -\frac{1}{9 \tilde V_t} \left(\sum\limits_{I\in t} \tilde A_I \tilde u^{(1I)}_i n^I_i\right) -\frac{1}{9\tilde V_t} \nabla_{\tilde X_j} \tilde u^{(0I)}_i \left(\tilde x^{I\!J}_j n^J_i \tilde A_J + \tilde x^{I\!K}_j n^K_i \tilde A_K + \tilde x^{I\!L}_j n^L_i \tilde A_L \right) \label{eq:ev0}  \\
\tilde\varepsilon^{(1)}_{\tilde V} &= -\frac{1}{9 \tilde V_t} \left(\sum\limits_{I\in t} \tilde A_I \tilde u^{(2I)}_i n^I_i\right) -\frac{1}{9\tilde V_t} \nabla_{\tilde X_j} \tilde u^{(1I)}_i \left(\tilde x^{I\!J}_j n^J_i \tilde A_J + \tilde x^{I\!K}_j n^K_i \tilde A_K + \tilde x^{I\!L}_j n^L_i \tilde A_L \right)  \nonumber \\
 &-\frac{1}{9\tilde V_t} \nabla_{\tilde X_j} \nabla_{\tilde X_k} \tilde u^{(0I)}_i \left(\tilde x^{I\!J}_j \tilde x^{I\!J}_k n^J_i \tilde A_J + \tilde x^{I\!K}_j \tilde x^{I\!K}_k n^K_i \tilde A_K + \tilde x^{I\!L}_j \tilde x^{I\!L}_k n^L_i \tilde A_L \right)  \label{eq:ev1} 
\end{align}
\end{subequations}
The term $\tilde\varepsilon^{(1)}_V$ can be simplified considering that the normals of the tetrahedral faces can be expressed as the cross product of its two edges 
\begin{align}
n^J_i &= \frac{1}{2 \tilde A_J}\levicivita_{ikl} \tilde x^{I\!K}_k \tilde x^{I\!L}_l &
n^K_i &= \frac{1}{2 \tilde A_K}\levicivita_{ikl} \tilde x^{I\!L}_k \tilde x^{I\!J}_l &
n^L_i &= \frac{1}{2 \tilde A_L}\levicivita_{ikl} \tilde x^{I\!J}_k \tilde x^{I\!K}_l
\end{align}
and volume of the tetrahedron is given by any (positive) permutation of the following expression
\begin{align}
V_t &= -\frac{1}{6}\levicivita_{ikl} x^{I\!J}_i x^{I\!K}_k x^{I\!L}_l
\end{align}
Expression~\eqref{eq:ev0}  then yields
\begin{align}
\tilde\varepsilon^{(0)}_V &= -\frac{1}{9 \tilde V_t} \left(\sum\limits_{I\in t} \tilde A_I \tilde u^{(1I)}_i n^I_i\right) + \frac{1}{3} \mathrm{tr} \left( \nabla_{\tilde X_j} \tilde u^{(0I)}_i \right) \label{eq:ev0X}
\end{align}

\section*{Data availability}
Computational models, scripts to run them, and their results are available at Zenodo repository under DOI \href{https://doi.org/10.5281/zenodo.15836839}{10.5281/zenodo.15836839}.

\section*{Acknowledgement}
The first author acknowledges financial support from the Czech Science Foundation under project number 24-11845S. The second author acknowledges support from the Engineering Research and Development Center (ERDC), Construction Engineering Research Laboratory (CERL), under contract no. W9132T22C0015.  We also thank Samuel Forest from Mines ParisTech for valuable comments on his outstanding paper~\parencite{ForPra-01} and Milan Jirásek, Martin Doškář, Jan Zeman and Martin Horák from Czech Technical University for encouraging and stimulating discussion.

\printbibliography

@article{SanMun25,
author = {Santos, Augusto and Muñoz-Rojas, Pablo},
year = {2025},
pages = {},
title = {A modified novel implementation of asymptotic homogenization {(NIAH)} to model frame-like periodic materials},
journal = {Meccanica},
issn = {1572-9648},
doi = {10.1007/s11012-025-02029-8}
}

@article{PerShm18,
title = {Dynamic homogenization of composite and locally resonant flexural systems},
journal = {Journal of the Mechanics and Physics of Solids},
volume = {119},
pages = {43--59},
year = {2018},
issn = {0022-5096},
doi = {10.1016/j.jmps.2018.06.011},
author = {René Pernas-Salomón and Gal Shmuel},
}

@article{TorFan-25,
title = {A multiscale approach to visco-electro-elastic complex materials: {Asymptotic} homogenization versus high-frequency continualization schemes},
journal = {International Journal of Engineering Science},
volume = {216},
pages = {104331},
year = {2025},
issn = {0020-7225},
doi = {10.1016/j.ijengsci.2025.104331},
author = {Rosaria {Del Toro} and Francesca Fantoni and Maria Laura {De Bellis} and Andrea Bacigalupo},
}

@article{BacBad-25,
title = {Enhanced high-frequency continualization scheme for inertial beam-lattice metamaterials},
journal = {International Journal of Mechanical Sciences},
volume = {286},
pages = {109794},
year = {2025},
issn = {0020-7403},
doi = {10.1016/j.ijmecsci.2024.109794},
author = {Andrea Bacigalupo and Paolo Badino and Vito Diana and Luigi Gambarotta},
}

@article{TorFan-24,
author = {Rosaria {Del Toro} and Francesca Fantoni and Maria Laura {De Bellis} and Andrea Bacigalupo},
title = {Dynamic continualization of mechanical metamaterials with quasi-periodic microstructure},
journal = {Phil. Trans. R. Soc. A},
volume = {382},
issue = {2278},
pages = {20230353},
year = {2024},
doi = {10.1098/rsta.2023.0353},
}

@article{CusZho14,
author = {Gianluca Cusatis  and Xinwei Zhou},
title = {High-Order Microplane Theory for Quasi-Brittle Materials with Multiple Characteristic Lengths},
journal = {Journal of Engineering Mechanics},
volume = {140},
number = {7},
pages = {04014046},
year = {2014},
doi = {10.1061/(ASCE)EM.1943-7889.0000747},
}

@article{TorBel-23,
title = {High frequency multi-field continualization scheme for layered magneto-electro-elastic materials},
journal = {International Journal of Solids and Structures},
volume = {282},
pages = {112431},
year = {2023},
issn = {0020-7683},
doi = {10.1016/j.ijsolstr.2023.112431},
author = {Rosaria {Del Toro} and Maria Laura {De Bellis} and Andrea Bacigalupo},
}

@article{GaoHe-25,
title = {A dynamic homogenization method for elastic wave band gap and initial-boundary value problem analysis of piezoelectric composites with elastic and viscoelastic periodic layers},
journal = {Journal of the Mechanics and Physics of Solids},
volume = {197},
pages = {106048},
year = {2025},
issn = {0022-5096},
doi = {10.1016/j.jmps.2025.106048},
author = {Mengyuan Gao and Zhelong He and Jie Liu and Chaofeng Lü and Guannan Wang},
}

@article{MuhLim-19,
title = {Built-up structural steel sections as seismic metamaterials for surface wave attenuation with low frequency wide bandgap in layered soil medium},
journal = {Engineering Structures},
volume = {188},
pages = {440-451},
year = {2019},
issn = {0141-0296},
doi = {10.1016/j.engstruct.2019.03.046},
author = {Muhammad and C.W. Lim and J.N. Reddy},
}

@phdthesis{PradelThesis,
  title        = {Homogénéisation des milieux discrets périodiques orientés. Une application aux mousses},
  author       = {Francis Pradel},
  year         = 1998,
  month        = {12},
  note         = {in French},
  school       = {Ecole Nationale des Ponts et Chaussées},
  type         = {PhD thesis}
}

@article{EliVor-20,
author = {Jan Eli\'{a}\v{s} and Miroslav Vo\v{r}echovsk\'{y} and V\'{a}clav Sad\'{i}lek},
title = {Periodic version of the minimax distance criterion for Monte Carlo integration},
journal = {Advances in Engineering Software},
year = {2020},
volume = {149},
pages = {102900},
doi = {10.1016/j.advengsoft.2020.102900},
issn = {0965-9978}
}

@article{SchCus12,
author = {Edward A. Schauffert  and Gianluca Cusatis },
title = {Lattice Discrete Particle Model for Fiber-Reinforced Concrete. {I}: {T}heory},
journal = {Journal of Engineering Mechanics},
volume = {138},
number = {7},
pages = {826--833},
year = {2012},
doi = {10.1061/(ASCE)EM.1943-7889.0000387},
}

@article{BarVar01,
title = {The asymmetry of stress in granular media},
journal = {International Journal of Solids and Structures},
volume = {38},
number = {2},
pages = {353--367},
year = {2001},
issn = {0020-7683},
doi = {10.1016/S0020-7683(00)00021-4},
author = {J.P. Bardet and I. Vardoulakis},
}

@article{ChaKuh05,
title = {On virtual work and stress in granular media},
journal = {International Journal of Solids and Structures},
volume = {42},
number = {13},
pages = {3773--3793},
year = {2005},
issn = {0020-7683},
doi = {10.1016/j.ijsolstr.2004.11.011},
author = {Ching S. Chang and Matthew R. Kuhn},
}

@article{AldHav-22,
title = {Effect of creep on corrosion-induced cracking},
journal = {Engineering Fracture Mechanics},
volume = {264},
pages = {108310},
year = {2022},
issn = {0013-7944},
doi = {10.1016/j.engfracmech.2022.108310},
author = {Ismail Aldellaa and Petr Havlásek and Milan Jirásek and Peter Grassl},
}

@article{AthWhe-18,
title = {Hydro-mechanical network modelling of particulate composites},
journal = {International Journal of Solids and Structures},
volume = {130-131},
pages = {49-60},
year = {2018},
issn = {0020-7683},
doi = {10.1016/j.ijsolstr.2017.10.017},
author = {Ignatios Athanasiadis and Simon J. Wheeler and Peter Grassl},
}

@article{TroLal-25,
title = {Lattice discrete particle model simulations of energetic size effect and its implications for shear design specifications of reinforced concrete squat walls},
journal = {Engineering Structures},
volume = {322},
pages = {119085},
year = {2025},
issn = {0141-0296},
doi = {10.1016/j.engstruct.2024.119085},
author = {Matthew Troemner and Erol Lale and Gianluca Cusatis},
}

@article{YinTro-24,
title = {An interprocess communication-based two-way coupling approach for implicit–explicit multiphysics lattice discrete particle model simulations},
journal = {Engineering Fracture Mechanics},
volume = {310},
pages = {110515},
year = {2024},
issn = {0013-7944},
doi = {10.1016/j.engfracmech.2024.110515},
author = {Hao Yin and Matthew Troemner and Weixin Li and Erol Lale and Lifu Yang and Lei Shen and Mohammed Alnaggar and Giovanni {Di Luzio} and Gianluca Cusatis},
}

@article{PatZha-23,
title = {Confinement and alkali-silica reaction in concrete: Review and numerical investigation},
journal = {International Journal of Solids and Structures},
volume = {277-278},
pages = {112341},
year = {2023},
issn = {0020-7683},
doi = {10.1016/j.ijsolstr.2023.112341},
author = {Madura Pathirage and Boqin Zhang and Mohammed Alnaggar and Gianluca Cusatis},
}

@article{PatTon-23,
title = {Discrete modeling of concrete failure and size-effect},
journal = {Theoretical and Applied Fracture Mechanics},
volume = {124},
pages = {103738},
year = {2023},
issn = {0167-8442},
doi = {10.1016/j.tafmec.2022.103738},
author = {Madura Pathirage and Danyang Tong and Flavien Thierry and Gianluca Cusatis and David Grégoire and Gilles Pijaudier-Cabot},
}

@article{GomBha-21,
author = {Shady Gomaa  and Tathagata Bhaduri  and Mohammed Alnaggar },
title = {Coupled Experimental and Computational Investigation of the Interplay between Discrete and Continuous Reinforcement in Ultrahigh Performance Concrete Beams. I: Experimental Testing},
journal = {Journal of Engineering Mechanics},
volume = {147},
number = {9},
pages = {04021049},
year = {2021},
doi = {10.1061/(ASCE)EM.1943-7889.0001948},
}

@article{JinBur-16,
title = {Lattice discrete particle modeling of fiber reinforced concrete: Experiments and simulations},
journal = {European Journal of Mechanics - A/Solids},
volume = {57},
pages = {85-107},
year = {2016},
issn = {0997-7538},
doi = {10.1016/j.euromechsol.2015.12.002},
author = {Congrui Jin and Nicola Buratti and Marco Stacchini and Marco Savoia and Gianluca Cusatis},
}

@article{Kaw78,
title = {New discrete models and their application to seismic response analysis of structures},
journal = {Nuclear Engineering and Design},
volume = {48},
number = {1},
pages = {207-229},
year = {1978},
issn = {0029-5493},
doi = {10.1016/0029-5493(78)90217-0},
author = {Tadahiko Kawai},
}

@article{BolXu-25,
title = {Does printing direction influence the bond between 3D printed polymeric reinforcement and cementitious matrix?},
journal = {Engineering Failure Analysis},
volume = {174},
pages = {109471},
year = {2025},
issn = {1350-6307},
doi = {10.1016/j.engfailanal.2025.109471},
author = {Rowin J.M. Bol and Yading Xu and Mladena Luković and Branko Šavija},
}

@article{BolSai98,
title = {Fracture analyses using spring networks with random geometry},
journal = {Engineering Fracture Mechanics},
volume = {61},
number = {5},
pages = {569-591},
year = {1998},
issn = {0013-7944},
doi = {10.1016/S0013-7944(98)00069-1},
author = {J.E. Bolander and S. Saito},
}

@article{JosKun-23,
title = {Reproducible estimations of internal corrosion distribution from surface cracks using MPC-RBSM},
journal = {Engineering Fracture Mechanics},
volume = {292},
pages = {109642},
year = {2023},
issn = {0013-7944},
doi = {10.1016/j.engfracmech.2023.109642},
author = {Suhas S. Joshi and Vikas Singh Kuntal and John E. Bolander and Kohei Nagai},
}

@article{ShoNak-24,
title = {Mesoscopic evaluation of the bond behavior of concrete with deformed rebar subjected to passive confinement employing 3D discrete model},
journal = {Engineering Fracture Mechanics},
volume = {295},
pages = {109790},
year = {2024},
issn = {0013-7944},
doi = {10.1016/j.engfracmech.2023.109790},
author = {Muhammad {Shoaib Karam} and Hikaru Nakamura and Yoshihito Yamamoto and Muhammad Tahir and Rashid Hameed},
}

@article{YinJia-24,
title = {Mesoscale discrete simulation of flexural behavior of FRP-strengthened RC beams using 3D RBSM},
journal = {Engineering Structures},
volume = {310},
pages = {118131},
year = {2024},
issn = {0141-0296},
doi = {10.1016/j.engstruct.2024.118131},
author = {Wenliang Yin and Cheng Jiang and Kohei Nagai}
}

@article{ParCho-24,
title = {Investigation of high strain rate effects on strain-hardening cementitious composites using Voronoi-cell lattice models},
journal = {Cement and Concrete Composites},
volume = {147},
pages = {105408},
year = {2024},
issn = {0958-9465},
doi = {10.1016/j.cemconcomp.2023.105408},
author = {Ji Woon Park and Bonhwi Choo and John E. Bolander and Yun Mook Lim},
}

@article{SchMie92,
title = {Simple lattice model for numerical simulation of fracture of concrete materials and structures},
journal = {Materials and Structures},
volume = {25},
issue = {9},
pages = {534--542},
year = {1992},
issn = {1871-6873},
doi = {10.1007/BF02472449},
author = {Schlangen, E. and {van Mier}, J. G. M.},
}

@article{EliSta12,
author = {Jan Eli\'{a}\v{s} and Henrik Stang},
title = {Lattice Modeling of Aggregate Interlocking in Concrete},
journal = {International Journal of Fracture},
year = {2012},
volume = {175},
pages = {1--11},
doi = {10.1007/s10704-012-9677-3},
issn = {0376-9429}
}

@article{LyuPat-23,
title = {Dissipation mechanisms of crack-parallel stress effects on fracture process zone in concrete},
journal = {Journal of the Mechanics and Physics of Solids},
volume = {181},
pages = {105439},
year = {2023},
issn = {0022-5096},
doi = {10.1016/j.jmps.2023.105439},
author = {Yuhui Lyu and Madura Pathirage and Hoang T. Nguyen and Zdeněk P. Bažant and Gianluca Cusatis},
}

@article{ZhaWan-05,
author = {Zhang, Hongwu and Wang, Huia nd Liu, Guozhen},
title = {Quadrilateral isoparametric finite elements for plane elastic {C}osserat bodies},
journal = {Acta Mechanica Sinica},
volume = {21},
number = {4},
pages = {388--394},
doi = {10.1007/s10409-005-0041-y},
year = {2005}
}

@article{ChuHul93,
author = {Chung, J. and Hulbert, G. M.},
title = "A Time Integration Algorithm for Structural Dynamics With Improved Numerical Dissipation: The Generalized-{$\alpha$} Method",
journal = {Journal of Applied Mechanics},
volume = {60},
number = {2},
pages = {371--375},
year = {1993},
issn = {0021-8936},
doi = {10.1115/1.2900803},
}

@article{EliVor16,
author = {Jan Eli\'{a}\v{s} and Miroslav Vo\v{r}echovsk\'{y}},
title = {Modification of the Audze–Eglajs criterion to achieve a uniform distribution of sampling points},
journal = {Advances in Engineering Software},
year = {2016},
volume = {100},
pages = {82--96},
doi = {10.1016/j.advengsoft.2016.07.004},
issn = {0965-9978}
}

@article{EliCus25,
author = {Jan Eli\'{a}\v{s} and Gianluca Cusatis},
title = {Macroscopic stress, couple stress and flux tensors derived through energetic equivalence from microscopic continuous and discrete heterogeneous finite representative volumes},
journal = {Computer Methods in Applied Mechanics and Engineering},
volume = {436},
pages = {117688},
year = {2025},
doi = {10.1016/j.cma.2024.117688},
issn = {0045-7825}
}

@article{ForPra-01,
title = {Asymptotic analysis of heterogeneous Cosserat media},
journal = {International Journal of Solids and Structures},
volume = {38},
number = {26},
pages = {4585--4608},
year = {2001},
issn = {0020-7683},
doi = {10.1016/S0020-7683(00)00295-X},
author = {Samuel Forest and Francis Pradel and Karam Sab},
}

@book{Vard18_book,
  title     = "Cosserat Continuum Mechanics",
  author    = "Ioannis Vardoulakis",
  year      = 2019,
  publisher = "Springer International Publishing",
  doi   = "10.1007/978-3-319-95156-0",
  isbn = "978-3-319-95155-3",
  note = "With Applications to Granular Media"
}

@article{EliYin-22,
title = "Homogenization of discrete diffusion models by asymptotic expansion",
journal = "International Journal for Numerical and Analytical Methods in Geomechanics",
volume = "",
number = "",
pages = "",
note = "under review",
year = "2022",
issn = "0363-9061",
doi = "",
author = "Jan Eliáš and Hao Yin and Gianluca Cusatis",
}

@article{EliCus22,
author = {Jan Eli\'{a}\v{s} and Gianluca Cusatis},
title = {Homogenization of discrete mesoscale model of concrete for coupled mass transport and mechanics by asymptotic expansion},
journal = {Journal of the Mechanics and Physics of Solids},
year = {2022},
volume = {167},
pages = {105010},
doi = {10.1016/j.jmps.2022.105010},
issn = {0022-5096}
}

@article{BolEli-21,
	author = {John E. Bolander and Jan Eli\'{a}\v{s} and Gianluca Cusatis and Kohei Nagai},
	title = {Discrete mechanical models of concrete fracture},
	journal = {Engineering Fracture Mechanics},
	year = {2021},
	volume = {257},
	pages = {108030},
	doi = {10.1016/j.engfracmech.2021.108030},
	issn = {0013-7944}
}

@article{GitAsk-07,
	title = {Representative volume: Existence and size determination},
	journal = {Engineering Fracture Mechanics},
	volume = {74},
	number = {16},
	pages = {2518--2534},
	year = {2007},
	issn = {0013-7944},
	doi = {10.1016/j.engfracmech.2006.12.021},
	author = {I.M. Gitman and H. Askes and L.J. Sluys},
}

@article{Ung13,
	title = {An {FE2-X1} approach for multiscale localization phenomena},
	journal = {Journal of the Mechanics and Physics of Solids},
	volume = {61},
	number = {4},
	pages = {928--948},
	year = {2013},
	issn = {0022-5096},
	doi = {10.1016/j.jmps.2012.12.010},
	author = {Jörg F. Unger},
}

@article{RezCus16,
title = "Asymptotic expansion homogenization of discrete fine-scale models with rotational degrees of freedom for the simulation of quasi-brittle materials",
journal = "Journal of the Mechanics and Physics of Solids",
volume = "88",
pages = "320--345",
year = "2016",
issn = "0022-5096",
doi = "10.1016/j.jmps.2016.01.001",
author = "Roozbeh Rezakhani and Gianluca Cusatis"
}

@article{FisChe-07,
title = "Generalized mathematical homogenization of atomistic media at finite temperatures in three dimensions",
journal = "Computer Methods in Applied Mechanics and Engineering",
volume = "196",
number = "4",
pages = "908--922",
year = "2007",
issn = "0045-7825",
doi = "10.1016/j.cma.2006.08.001",
author = "Jacob Fish and Wen Chen and Renge Li"
}

@article{Eli17,
author = {Jan Eli\'{a}\v{s}},
title = {Boundary Layer Effect on Behavior of Discrete Models},
journal = {Materials},
year = {2017},
volume = {10},
pages = {157},
doi = {10.3390/ma10020157},
issn = {1996-1944}
}

@article{ChaLia-90,
	title = {Constitutive relation for a particulate medium with the effect of particle rotation},
	journal = {International Journal of Solids and Structures},
	volume = {26},
	number = {4},
	pages = {437--453},
	year = {1990},
	issn = {0020-7683},
	doi = {10.1016/0020-7683(90)90067-6},
	author = {Ching S. Chang and Ching L. Liao},
}

@InProceedings{RotSel81,
author = {Rothenburg, Leo and Selvadurai, Patrick},
year = {1981},
pages = {469--486},
title = {A Micromechanical Definition of the (C)auchy Stress Tensor for Particulate Media},
booktitle = {Proc. Int. Symp. Mechanical Behavior of Structured Media},
location = "Ottawa",
}

@article{NicHad-13,
title = {On the definition of the stress tensor in granular media},
journal = {International Journal of Solids and Structures},
volume = {50},
number = {14},
pages = {2508--2517},
year = {2013},
issn = {0020-7683},
doi = {10.1016/j.ijsolstr.2013.04.001},
author = {François Nicot and Nejib Hadda and Mohamed Guessasma and Jerome Fortin and Olivier Millet},
}

@article{Web66,
	author = {Weber, J.},
	title = "{Recherches concernant les contraintes intergranulaires dans les
	milieux pulvérulents}",
	journal = {Bulletin de Liaison des Ponts-et-Chaussées},
	volume = {20},
	pages = {1--20},
	year = {1966},
}

@book{Love1927,
	title     = "A treatise on the mathematical theory of elasticity",
	author    = "Augustus E.H. Love",
	year      = 1927,
	publisher = "Cambridge University Press",
	volume = "1",	
}

@article{YanReg19,
title = {Definition and symmetry of averaged stress tensor in granular media and its {3D} {DEM} inspection under static and dynamic conditions},
journal = {International Journal of Solids and Structures},
volume = {161},
pages = {243--266},
year = {2019},
issn = {0020-7683},
doi = {10.1016/j.ijsolstr.2018.11.021},
author = {Beichuan Yan and Richard A. Regueiro},
}

@article{LinWu16,
title = {Asymmetry of the stress tenor in granular materials},
journal = {Powder Technology},
volume = {293},
pages = {113--120},
year = {2016},
issn = {0032-5910},
doi = {10.1016/j.powtec.2015.11.034},
author = {Jia Lin and Wei Wu},
}

@article{ChrMeh-81,
    author = {Christoffersen, J. and Mehrabadi, M. M. and Nemat-Nasser, S.},
    title = {A Micromechanical Description of Granular Material Behavior},
    journal = {Journal of Applied Mechanics},
    volume = {48},
    number = {2},
    pages = {339--344},
    year = {1981},
    issn = {0021-8936},
    doi = {10.1115/1.3157619},
}

@article{BalMar07,
    author = {R. Balevičius and D. Markauskas},
    title = {Numerical stress analysis of granular material},
    journal = {Mechanics},
    volume = {66},
    number = {4},
    pages = {12-17},
    year = {2007},
}

@article{RezZho-17,
	title = {Adaptive multiscale homogenization of the lattice discrete particle model for the analysis of damage and fracture in concrete},
	journal = {International Journal of Solids and Structures},
	volume = {125},
	pages = {50-67},
	year = {2017},
	issn = {0020-7683},
	doi = {h10.1016/j.ijsolstr.2017.07.016},
	author = {Roozbeh Rezakhani and Xinwei Zhou and Gianluca Cusatis}
}

@article{EliVor20,
	author = {Jan Eli\'{a}\v{s} and Miroslav Vo\v{r}echovsk\'{y}},
	title = {Fracture in random quasibrittle media: {I}. {D}iscrete mesoscale simulations of load capacity and fracture process zone},
	journal = {Engineering Fracture Mechanics},
	year = {2020},
	volume = {235},
	pages = {107160},
	doi = {10.1016/j.engfracmech.2020.107160},
	issn = {0013-7944}
}

@article{Eli20,
	author = {Jan Eli\'{a}\v{s}},
	title = {Elastic properties of isotropic discrete systems: Connections between geometric structure and Poisson’s ratio},
	journal = {International Journal of Solids and Structures},
	year = {2020},
	volume = {191-192},
	pages = {254--263},
	doi = {10.1016/j.ijsolstr.2019.12.012},
	issn = {0020-7683}
}

@article{CusPel-11,
	title = {Lattice Discrete Particle Model ({LDPM}) for failure behavior of concrete. {I}: Theory},
	journal = {Cement and Concrete Composites},
	volume = {33},
	number = {9},
	pages = {881--890},
	year = {2011},
	issn = {0958-9465},
	doi = {10.1016/j.cemconcomp.2011.02.011},
	author = {Gianluca Cusatis and Daniele Pelessone and Andrea Mencarelli},
}

@article{CusMen-11,
	title={Lattice discrete particle model ({LDPM}) for failure behavior of concrete. {II}: Calibration and validation},
	author={Cusatis, Gianluca and Mencarelli, Andrea and Pelessone, Daniele and Baylot, James},
	journal={Cement and Concrete composites},
	volume={33},
	number={9},
	pages={891--905},
	year={2011}
}

@article{RezAln-19,
	title = {Multiscale Homogenization Analysis of Alkali–Silica Reaction ({ASR}) Effect in Concrete},
	journal = {Engineering},
	volume = {5},
	number = {6},
	pages = {1139--1154},
	year = {2019},
	issn = {2095-8099},
	doi = {10.1016/j.eng.2019.02.007},
	author = {Roozbeh Rezakhani and Mohammed Alnaggar and Gianluca Cusatis},
}

@article{JiaBri-24,
author = {Jia, Dongge and Brigham, John C. and Fascetti, Alessandro},
title = {An efficient static solver for the lattice discrete particle model},
journal = {Computer-Aided Civil and Infrastructure Engineering},
volume = {39},
number = {23},
pages = {3531-3551},
doi = {10.1111/mice.13306},
year = {2024}
}

@book{San80,
  title={Non-Homogeneous Media and Vibration Theory},
  author={Sanchez-Palencia, Enrique},
  isbn={9780387100005},
  series={Lecture Notes in Physics},
  year={1980},
  publisher={Springer-Verlag: Berlin/Heidelberg},
  address   = {Germany; New York, NY, USA},
}

@article{ZhuJia-25,
author = {Zhu, Yingbo and Jia, Dongge and Brigham, John C. and Fascetti, Alessandro},
title = {Coupled lattice discrete particle model for the simulation of water and chloride transport in cracked concrete members},
year = {2025},
volume = {40},
number = {8},
issn = {1093-9687},
doi = {10.1111/mice.13385},
journal = {Computer-Aided Civil and Infrastructure Engineering},
pages = {982–1003},
}

@article{CusCed07,
	title = {Two-scale study of concrete fracturing behavior},
	journal = {Engineering Fracture Mechanics},
	volume = {74},
	number = {1},
	pages = {3--17},
	year = {2007},
	note = {Fracture of Concrete Materials and Structures},
	issn = {0013-7944},
	doi = {10.1016/j.engfracmech.2006.01.021},
	author = {Gianluca Cusatis and Luigi Cedolin},
}

@article{Eli16,
	author = {Jan Eli\'{a}\v{s}},
	title = {Adaptive technique for discrete models of fracture},
	journal = {International Journal of Solids and Structures},
	year = {2016},
	volume = {100--101},
	pages = {376--387},
	doi = {10.1016/j.ijsolstr.2016.09.008},
	issn = {0020-7683}
}

@book{BenLio-78,
	author = {Bensoussan, A. and Lions, J.-L. and Papanicolaou, G.},
	publisher = {North-Holland},
	title = {Asymptotic Analysis for Periodic Structures},
	year = 1978
}

@article{SmiBre-98,
	title = {Prediction of the mechanical behavior of nonlinear heterogeneous systems by multi-level finite element modeling},
	journal = {Computer Methods in Applied Mechanics and Engineering},
	volume = {155},
	number = {1},
	pages = {181--192},
	year = {1998},
	issn = {0045-7825},
	doi = {10.1016/S0045-7825(97)00139-4},
	author = {R.J.M. Smit and W.A.M. Brekelmans and H.E.H. Meijer},
}

@article{TurHoo-18,
	title = {Multiscale analysis of mixed-mode fracture and effective traction-separation relations for composite materials},
	journal = {Journal of the Mechanics and Physics of Solids},
	volume = {117},
	pages = {88--109},
	year = {2018},
	issn = {0022-5096},
	doi = {10.1016/j.jmps.2018.04.009},
	author = {Sergio Turteltaub and Niels {van Hoorn} and Wim Westbroek and Christian Hirsch},
}

@article{OliCai-15,
	title = {Continuum approach to computational multiscale modeling of propagating fracture},
	journal = {Computer Methods in Applied Mechanics and Engineering},
	volume = {294},
	pages = {384--427},
	year = {2015},
	issn = {0045-7825},
	doi = {10.1016/j.cma.2015.05.012},
	author = {J. Oliver and M. Caicedo and E. Roubin and A.E. Huespe and J.A. Hernández},
}

@article{TanDon-19,
	title = {Simulation of strain localization with discrete element-{C}osserat continuum finite element two scale method for granular materials},
	journal = {Journal of the Mechanics and Physics of Solids},
	volume = {122},
	pages = {450-471},
	year = {2019},
	issn = {0022-5096},
	doi = {10.1016/j.jmps.2018.09.029},
	author = {Hongxiang Tang and Yan Dong and Ting Wang and Yifeng Dong},
}

@article{LebPon-21,
	title = {Effective toughness of disordered brittle solids: {A} homogenization framework},
	journal = {Journal of the Mechanics and Physics of Solids},
	volume = {153},
	pages = {104463},
	year = {2021},
	issn = {0022-5096},
	doi = {10.1016/j.jmps.2021.104463},
	author = {Mathias Lebihain and Laurent Ponson and Djimédo Kondo and Jean-Baptiste Leblond},
}

@article{AlaGan-21,
	title = {Construction of micromorphic continua by homogenization based on variational principles},
	journal = {Journal of the Mechanics and Physics of Solids},
	volume = {153},
	pages = {104278},
	year = {2021},
	issn = {0022-5096},
	doi = {10.1016/j.jmps.2020.104278},
	author = {S.E. Alavi and J.F. Ganghoffer and H. Reda and M. Sadighi},
}

@book{AurBou-09,
	author = {Jean-Louis Auriault and Claude Boutin and Christian Geindreau},
	publisher = {John Wiley \& Sons, Ltd},
	isbn = {9780470612033},
	title = {Homogenization of Coupled Phenomena in Heterogenous Media},
	pages = {1-476},
	doi = {10.1002/9780470612033},
	year = {2009},
}

@article{Hre-41,
    author = {Hrennikoff, A.},
    title = {Solution of Problems of Elasticity by the Framework Method},
    journal = {Journal of Applied Mechanics},
    volume = {8},
    number = {4},
    pages = {A169--A175},
    year = {1941},
    issn = {0021-8936},
    doi = {10.1115/1.4009129},
}

@article{CunStr79,
author = {Cundall, P. A. and Strack, O. D. L.},
title = {A discrete numerical model for granular assemblies},
journal = {Géotechnique},
volume = {29},
number = {1},
pages = {47-65},
year = {1979},
doi = {10.1680/geot.1979.29.1.47},
}

@inproceedings{Cun71,
    author = {Cundall, P. A.},
    title = {A computer model for simulating  progressive large scale movements in blocky rock systems},
    volume = {Proc., Int. Symp. Rock Fracture, ISRM},
    year = {1971},
    address = {Nancy, Franc},
    pages = {2--8},
}

@techreport{CunStr78,
  author      = "Cundall, P. A. and Strack, O. D. L.",
  title       = "{BALL-A} program to model granular media using distinct element method",
  institution = "Advanced Tech. Group, Dames and Moore, London, U.K.",
  year        = "1978"
}

@article{JivEng-13,
title = "Pore space and brittle damage evolution in concrete",
journal = "Engineering Fracture Mechanics",
volume = "110",
number = "0",
pages = "378--395",
year = "2013",
issn = "0013-7944",
doi = "10.1016/j.engfracmech.2013.05.007",
author = "Andrey P. Jivkov and Dirk L. Engelberg and Robert Stein and Mihail Petkovski"
}

@ARTICLE{ManMie11,
author={Hau-Kit Man and Jan G.M. {van Mier}},
title={Damage distribution and size effect in numerical concrete from lattice analyses},
journal={Cement \& Concrete Composites},
year={2011},
volume={33},
number={9},
pages={867--880},
doi={10.1016/j.cemconcomp.2011.01.008},
issn = {0958-9465}
}

@article{ManMie08,
title = "Influence of particle density on {3D} size effects in the fracture of (numerical) concrete",
journal = "Mechanics of Materials",
volume = "40",
number = "6",
pages = "470--486",
year = "2008",
author = "Hau-Kit Man and Jan G.M. {van Mier}",
issn = "0167-6636",
doi = "10.1016/j.mechmat.2007.11.003"
}

@article{LukSav-16,
    author = {Lukovi\'{c}, Mladena and \v{S}avija, Branko and Schlangen, Erik and Ye, Guang and {van Breugel}, Klaas},
    title = {A {3D} lattice modelling study of drying shrinkage damage in concrete repair systems},
    journal = {Materials},
    volume = {9},
    pages = {575},
    doi = {10.3390/ma9070575},
    year = "2016",
    issn = {1996-1944},
}

@article{MunRag-13,
title = "Fracture studies on {3D} geometrically similar beams",
journal = "Engineering Fracture Mechanics",
volume = "98",
number = "0",
pages = "407--422",
year = "2013",
issn = "0013-7944",
doi = "10.1016/j.engfracmech.2012.11.012",
author = "Mahesh Mungule and B.K. Raghuprasad and S. Muralidhara"
}

@article{DonMag-99,
	author = {F. V. Donzé  and S.-A. Magnier  and L. Daudeville  and C. Mariotti  and L. Davenne },
	title = {Numerical Study of Compressive Behavior of Concrete at High Strain Rates},
	journal = {Journal of Engineering Mechanics},
	volume = {125},
	number = {10},
	pages = {1154--1163},
	year = {1999},
	doi = {10.1061/(ASCE)0733-9399(1999)125:10(1154)}
}

@article{EliVor-15,
    author = {Jan Eli\'{a}\v{s} and Miroslav Vo\v{r}echovsk\'{y} and Jan Sko\v{c}ek and Zden\v{e}k P. Ba\v{z}ant},
    title = {Stochastic discrete meso-scale simulations of concrete fracture: comparison to experimental data},
    journal = {Engineering Fracture Mechanics},
    year = {2015},
    volume = {135},
    pages = {1--16},
    doi = {10.1016/j.engfracmech.2015.01.004},
    issn = {0013-7944},
}

@article{CusBaz-03I,
author = {Gianluca Cusatis  and Zdeněk P. Bažant  and Luigi Cedolin },
title = {Confinement-Shear Lattice Model for Concrete Damage in Tension and Compression: {I}. Theory},
journal = {Journal of Engineering Mechanics},
volume = {129},
number = {12},
pages = {1439--1448},
year = {2003},
doi = {10.1061/(ASCE)0733-9399(2003)129:12(1439)}
}

@article{Zhu2022,
  author = {Zhu, Z. and Deng, B. and Huang, H. and Du, J.},
  title = {Modeling lattice metamaterials with deformable joints as an elastic micropolar continuum},
  journal = {AIP Advances},
  volume = {12},
  number = {6},
  pages = {065116},
  year = {2022},
  doi = {10.1063/5.0080503},
  url = {https://pubs.aip.org/aip/adv/article/12/6/065116/2819125/Modeling-lattice-metamaterials-with-deformable}
}

@article{Liu2015,
  author = {Liu, Y. and Zhang, X. and Zhang, S. and Liu, Z. and Zhang, X.},
  title = {Generalized Lorentz-Lorenz homogenization formulas for binary lattice metamaterials},
  journal = {Physical Review B},
  volume = {91},
  number = {20},
  pages = {205127},
  year = {2015},
  doi = {10.1103/PhysRevB.91.205127},
  url = {https://journals.aps.org/prb/abstract/10.1103/PhysRevB.91.205127}
}

@article{Plesha2021,
  author = {Plesha, E. and Alderson, K. and Evans, R. and Gatt, J. and Grima, J. and Miller, W. and Ravirala, N. and Smith, C. and Zied, K.},
  title = {Optimized lattice-based metamaterials for elastostatic cloaking},
  journal = {Proceedings of the Royal Society A: Mathematical, Physical and Engineering Sciences},
  volume = {477},
  number = {2254},
  pages = {20210418},
  year = {2021},
  doi = {10.1098/rspa.2021.0418},
  url = {https://royalsocietypublishing.org/doi/10.1098/rspa.2021.0418}
}

@article{Kadic2016,
  author = {Kadic, M. and Bückmann, T. and Schittny, R. and Wegener, M.},
  title = {Designing Perturbative Metamaterials from Discrete Models: From Veselago lenses to topological insulators},
  journal = {arXiv},
  year = {2016},
  eprint = {1612.02362},
  url = {https://arxiv.org/abs/1612.02362}
}

@article{chen2006generalized,
  title={A generalized space--time mathematical homogenization theory for bridging atomistic and continuum scales},
  author={Chen, Wen and Fish, Jacob},
  journal={International Journal for Numerical Methods in Engineering},
  volume={67},
  number={2},
  pages={253--271},
  year={2006},
  publisher={Wiley Online Library}
}
\end{document}